\documentclass[fleqn,usenatbib]{mnras}

\usepackage{newtxtext,newtxmath}
\usepackage{fancyhdr}

\usepackage[T1]{fontenc}
\usepackage{ae,aecompl}

\usepackage{graphicx}	            
\usepackage{amsmath}	            
\usepackage{caption}                
\usepackage{pdfpages}               
\usepackage{float}                  
\usepackage{longtable}
\usepackage{lscape}

\usepackage{amsfonts}
\usepackage{graphicx}
\usepackage{tikz}
\usetikzlibrary{shapes,arrows}

\tikzstyle{block} = [rectangle,thick, draw=black!100, 
    text width=2.4em, text centered, rounded corners, minimum height=1em,minimum width = 0.6em]
\tikzstyle{block2} = [rectangle,thick, draw=black!100, 
    text width=2.4em, text centered, rounded corners, minimum height=1em,minimum width = 1.2em]
\tikzstyle{line} = [draw,ultra thick, -latex']
\tikzstyle{curved line} = [draw, bend right=45,ultra thick, -latex']
\tikzstyle{dashed line} = [draw,ultra thick, -latex',dashed]
\tikzstyle{cloud} = [thick, ellipse, draw=black!100, 
    node distance=0.6cm, minimum width=2em,
    minimum height=1.2em] 
    
\title[Reconstructing the GMF using FRBs \& radio galaxies]{A method for reconstructing the Galactic magnetic field using dispersion of fast radio bursts and Faraday rotation of radio galaxies}

\author[Pandhi et al.]{
A. Pandhi$^{1,2}$,
S. Hutschenreuter$^{3}$,
J. L. West$^{2}$,
B. M. Gaensler$^{2,1}$, 
A. Stock$^{1,4}$
\\
$^{1}$David A. Dunlap Department of Astronomy and Astrophysics, University of Toronto, 50 St. George Street, Toronto, ON M5S 3H4, Canada\\
$^{2}$Dunlap Institute for Astronomy and Astrophysics, University of Toronto, 50 St. George Street, Toronto, ON M5S 3H4, Canada\\
$^{3}$Department of Astrophysics/IMAPP, Radboud University Nijmegen; P.O. Box 9010, 6500 GL Nijmegen, Netherlands\\
$^{4}$Canadian Institute for Theoretical Astrophysics, University of Toronto, 60 St. George Street, Toronto, ON, M5S 3H8, Canada\\
}

\begin{document}
\maketitle
\fancyhf{} 
\renewcommand{\headrulewidth}{0pt}
\thispagestyle{fancy}
\rhead{Compiled using MNRAS \LaTeX\ style file v3.0}

\begin{abstract}
With the rapid increase of fast radio burst (FRB) detections within the past few years, there is now a catalogue being developed for all-sky extragalactic dispersion measure (DM) observations in addition to the existing collection of all-sky extragalactic Faraday rotation measurements (RMs) of radio galaxies. We present a method of reconstructing all-sky information of the Galactic magnetic field component parallel to the line of sight, $B_{\parallel}$, using simulated observations of the RM and DM along lines of sight to radio galaxies and FRB populations, respectively. This technique is capable of distinguishing between different input Galactic magnetic field and thermal electron density models. Significant extragalactic contributions to the DM are the predominant impediment in accurately reconstructing the Galactic DM and $\left<B_{\parallel}\right>$ skies. We look at ways to improve the reconstruction by applying a filtering algorithm on the simulated DM lines of sight and we derive generalized corrections for DM observations at $|b|$ > 10~deg that help to disentangle Galactic and extragalactic DM contributions. Overall, we are able to reconstruct both large-scale Galactic structure and local features in the Milky Way's magnetic field from the assumed models. We discuss the application of this technique to future FRB observations and address possible differences between our simulated model and observed data, namely: adjusting the priors of the inference model, an unevenly distributed population of FRBs on the sky, and localized extragalactic DM structures.
\end{abstract}

\begin{keywords}
ISM: magnetic fields -- Galaxy: structure -- fast radio bursts
\end{keywords}

\section{Introduction} \label{intro}
The large-scale Galactic magnetic field (GMF) is important for understanding various astrophysical processes in the Galaxy, such as the structure and properties of the turbulent interstellar medium (ISM), molecular cloud collapse, star formation, cosmic-ray acceleration and propagation \citep[for reviews, see][and references therein]{2001RvMP...73.1031F, 2015ASSL..407..483H, 2019Galax...7...52J}. As such, modeling the GMF is an active field of research that utilizes a variety of observational tracers and analytical techniques to probe properties such as the orientation and magnitude of the magnetic field perpendicular, $B_{\perp}$, and parallel, $B_{\parallel}$, to the observer's line of sight (LOS). A summary of large-scale GMF tracers, their dependencies, pros, and cons is presented in Table 1 of \cite{2019Galax...7...52J}.

As linearly polarized emission propagates through a magneto-ionic medium, free electrons along the LOS induce Faraday rotation of the observed radio emission. This effect is characterized by a difference in the measured polarization angle between two observing frequencies and is proportional to the square of the observing wavelength, $\lambda^2$. The observed polarization position angle, $\chi(\lambda^2)$, from a background source with an intrinsic polarization angle, $\chi_0$, and Rotation Measure (RM) is described as:
\begin{equation}
\chi(\lambda^2) = \chi_0 + {\rm{RM}}\lambda^{2}\,. \label{eq:pa}
\end{equation}
RM is in turn related to the integrated number density of electrons and parallel component of the magnetic field along the LOS: 
\begin{equation}
{\rm{RM}}= 0.812\,{\rm rad}\,{\rm m}^{-2}\int_{0}^{d}\left[\frac{n_\text{e}(s)}{\text{cm}^{-3}}\right]\left[\frac{B_{\parallel}(s)}{\mu \text{G}}\right] \left(\frac{ds}{\text{pc}}\right)\label{eq:RM} \,,
\end{equation}
where $n_e$ is the thermal electron density, $d$ is the distance to the emitting source, and we define our limits of integration in accordance with \cite{2021MNRAS.507.4968F} such that we integrate from the source ($s=0$) to the observer ($s=d$).
Simultaneously, the emission is also dispersed by the free electrons. Empirically, this manifests as a wavelength dependent time delay of the observed emission which is parameterized by the dispersion measure (DM):
\begin{equation}
{\rm{DM}}= \int_{0}^{d}\left[\frac{n_\text{e}(s)}{\text{cm}^{-3}}\right] \left(\frac{ds}{\text{pc}}\right)\label{eq:DM} \,.
\end{equation}
Note that equations \ref{eq:RM} and \ref{eq:DM} assume a cosmological redshift of zero; later in Section \ref{simobs} we will explore RM and DM measurements for extragalactic sources and their redshift dependence. The ratio between RM and DM, along the same LOS, provides an estimate of the electron density weighted average magnetic field strength parallel to the LOS:
\begin{equation}
\left<B_{\parallel}\right> = \frac{\int_{0}^{d} n_{\rm{e}} B_{\parallel} ds}{\int_{0}^{d} n_{\rm{e}} ds} = 1.232\,\mu \text{G}\left(\frac{\text{RM}}{\text{rad}\, \text{m}^{-2}}\right)\left(\frac{\text{DM}}{\text{pc}\,\text{cm}^{-3}}\right)^{-1}\label{eq:B} \,.
\end{equation}
The mean direction of $\left<B_{\parallel}\right>$ is determined by positive (towards the observer) and negative (away from the observer) measurements of RM. The underlying assumption behind equation \ref{eq:B} is that the thermal electron density is uncorrelated with the LOS component of the magnetic field. However, this assumption may break down under certain circumstances, particularly at sub-kpc scales \citep[see][]{2003A&A...411...99B, 2021MNRAS.502.2220S}.

RMs of polarized radio sources, in conjunction with DM, are therefore an effective method for probing the strength and direction of the parallel component of the magnetic field, $\left<B_{\parallel}\right>$, in the intervening structure along the observed LOS. This technique has been widely used for sampling the 3D structure within the Galaxy using DMs and RMs of pulsars \citep[e.g.][]{1972ApJ...172...43M, 1974ApJ...188..637M, 1994MNRAS.268..497R, 1999MNRAS.306..371H, 2008MNRAS.386.1881N, 2019MNRAS.484.3646S, 2020MNRAS.496.2836N}. Observed Galactic pulsar RMs provide 3D information within the Galaxy to the extent that we know the distance to the pulsar but the overall sampling is limited (currently 1320 observed pulsars have measured RMs) \citep[\textit{ATNF} Pulsar Catalogue V1.66:][]{2005yCat.7245....0M}. Other studies have also created full-sky RM maps using polarized extragalactic radio sources \citep[e.g.][]{2012A&A...542A..93O, 2015A&A...575A.118O, 2020A&A...633A.150H, h21} without corresponding DM observations.The collection of extragalactic RMs is considerably larger than the sample of Galactic pulsar RMs, currently over 50000 sources\footnote{An up to date RM catalogue is at \href{https://github.com/CIRADA-Tools/RMTable}{https://github.com/CIRADA-Tools/RMTable}.}, primarily from observations of radio galaxies (\textcolor{blue}{Van Eck et al. in prep}), and is well sampled over almost all lines of sight (although it loses the 3D information since all observations probe the full extent of the Galactic structure along the LOS). In particular, \cite{2012A&A...542A..93O, 2015A&A...575A.118O} combined several extragalactic RM catalogues to reconstruct a full-sky map of observed total RM through the Galaxy, taking into account the uncertainties in the noise statistics. Following up on this work, \cite{2020A&A...633A.150H, h21} use an updated inference algorithm and incorporate the electron emission measure (EM) observed by the \textit{Planck} survey \citep{2016A&A...594A..10P, 2016A&A...596A.103P} to produce updated maps of the Galactic RM. Both Galactic and extragalactic RM data sets are expected to greatly improve as the Square Kilometre Array (SKA) and its associated pathfinder surveys come online  \citep[e.g.][]{2015aska.confE..40K, 2015aska.confE..92J, 2020Galax...8...53H}.

The study of DMs, however, has been predominantly limited to Galactic sources, namely pulsar observations (currently 3160 observed pulsars have DMs measured) \citep[\textit{ATNF} Pulsar Catalogue V1.66:][]{2005yCat.7245....0M}. Pulsars are transient sources on approximately millisecond to second timescales, allowing us to measure their DM. This is not the case for radio galaxies, which are continuously emitting at radio frequencies on these short timescales \citep{2016ApJ...823...93H}. Thus, there had not been a set of extragalactic DM data comparable in sky density to the extragalactic RM catalogue from radio galaxies, until the discovery of fast radio bursts (FRBs). FRBs are millisecond-duration radio transients that were first identified in 2007, when \cite{2007Sci...318..777L} detected a 30 Jy radio burst with a < 5 ms duration using data from the Parkes radio telescope. The observation and localization of FRBs has become a rapidly growing field, with the Canadian Hydrogen Intensity Mapping Experiment Fast Radio Burst (CHIME/FRB) Project recently releasing a catalogue of 536 FRBs \citep{2021ApJS..257...59A}. The inferred sky rate of FRB events at 1.4~GHz with a fluence above $3-4~\mathrm{Jy}~\mathrm{ms}$ is estimated between several hundred and a few thousand per sky per day \citep[e.g.][]{2013Sci...341...53T, 2016MNRAS.455.2207R, 2017AJ....154..117L}, implying that there is still a wealth of observational data to come in the future.

This expanding catalogue of FRBs\footnote{An up to date catalogue of FRB observations is at \href{https://herta-experiment.org/frbstats}{https://herta-experiment.org/frbstats}.} presents a large source of extragalactic DM observations that we cannot obtain using radio galaxies. While FRBs also provide RMs \citep{2021ApJ...920..138M}, the total number of observed FRBs is much smaller than the available RM sample from radio galaxies. Thus, FRB RMs will not be considered in this work. While there is no consensus as to the specific origin and physical mechanisms behind FRBs \citep[for a review, see][]{2019PhR...821....1P, 2019ARA&A..57..417C}, measurements of their DMs (typically hundreds of pc cm$^{-3}$) are too large to be entirely accredited to Milky Way (MW) contributions and are consistent with extragalactic origins. Additionally, several FRBs have been localized to other galaxies \citep[e.g.][]{2017Natur.541...58C, 2019Natur.572..352R, 2019Sci...365..565B, 2020Natur.577..190M}. 

In general, the observed DM from an extragalactic source, $\rm{DM}_{\rm{obs}}$ can be separated into distinct contributing components:
\begin{equation}
\rm{DM}_{\rm{obs}} = \rm{DM}_{\rm{disk}} + \rm{DM}_{\rm{halo}} + \rm{DM}_{\rm{IGM}} + \rm{DM}_{\rm{host}}\,, \label{eq:dm2}
\end{equation}
where $\rm{DM}_{\rm{disk}}$ is contribution from warm ionized gas in the MW disk ($T \lesssim 10^4$ K), $\rm{DM}_{\rm{halo}}$ is from the extended hot Galactic halo ($T \sim 10^6 - 10^7$ K), $\rm{DM}_{\rm{IGM}}$ is contributions from the intergalactic medium and intervening systems, and $\rm{DM}_{\rm{host}}$ is from the host galaxy of the source and its local environment \citep[and references therein]{2020ApJ...888..105Y}. Similarly, the $\rm{RM}_{\rm{obs}}$ is also broken down into a Galactic component (disk and halo) and extragalactic contributions, which are in general much smaller than the Galactic component: 
\begin{equation}
\rm{RM}_{\rm{obs}} = \rm{RM}_{\rm{disk}} + \rm{RM}_{\rm{halo}} + \rm{RM}_{\rm{xgal}}\,, \label{eq:rm2}
\end{equation}
Importantly, in equation \ref{eq:rm2}, the sum of all extragalactic RM contributions is folded into $\rm{RM}_{\rm{xgal}}$ rather than explicitly including $\rm{RM}_{\rm{IGM}}$ and $\rm{RM}_{\rm{host}}$ terms. This distinction is further examined in Section \ref{simobs}, which also includes a more detailed breakdown of the DM and RM contributions and estimates.

With the aforementioned growing FRB catalogue, the proposition of studying the full DM sky using methods similar to \cite{2012A&A...542A..93O, 2015A&A...575A.118O} and \cite{2020A&A...633A.150H} becomes increasingly plausible. With full-sky information of both RM and DM, one could, in principle, apply equation \ref{eq:B} to reconstruct an approximation of the $\left<B_{\parallel}\right>$ sky. In this work, we propose to establish and test a robust method of reconstructing $\left<B_{\parallel}\right>$ information across the sky using simulated data. Specifically, this will be done by assuming an underlying model of the GMF \citep[e.g.][]{2010RAA....10.1287S, 2012ApJ...757...14J, 2013MNRAS.431..683J, 2017A&A...600A..29T} and $n_e$ \citep[e.g.][]{2017ApJ...835...29Y, 2020ApJ...888..105Y}. Then, we simulate a distribution of FRB and radio galaxy populations with associated DM and RM contributions, respectively. By applying a reasonable lower limit to the observed flux from FRBs and determining the LOS and distance to these simulated points, we can recreate the conditions for typical extragalactic FRB and radio galaxy observations. Consequently, we use the simulated observations to build up approximations of the DM and RM skies with a joint inference algorithm and use these to reconstruct a map of the simulated $\left<B_{\parallel}\right>$ across the sky. By contrasting the resulting reconstruction with the underlying GMF model, we determine the validity and accuracy of the technique. Stemming from this analysis, we further explore methods of handling DM observations and correcting for the significant extragalactic contributions that naturally arise.

The remainder of this paper is structured as follows. In Section \ref{simobs} we discuss the details of generating observables from underlying assumptions of the GMF and $n_e$ distribution and simulating FRB and radio galaxy populations. Section \ref{inference} illustrates the use of Information Field Theory (IFT) \citep{2009PhRvD..80j5005E, 2019AnP...53100127E}, which incorporates methods from Bayesian statistics and statistical field theory, to infer underlying continuous fields from noisy incomplete data. The results of the joint inference and data reduction techniques utilized to minimize the extragalactic DM contributions are presented in Section \ref{results}. In Section \ref{discussion} we derive a correction factor to further reduce extragalactic DM and discuss the implications of our work.

\section{Simulated Observations} \label{simobs}

\begin{figure*}
    \centering
    \includegraphics[scale=0.690]{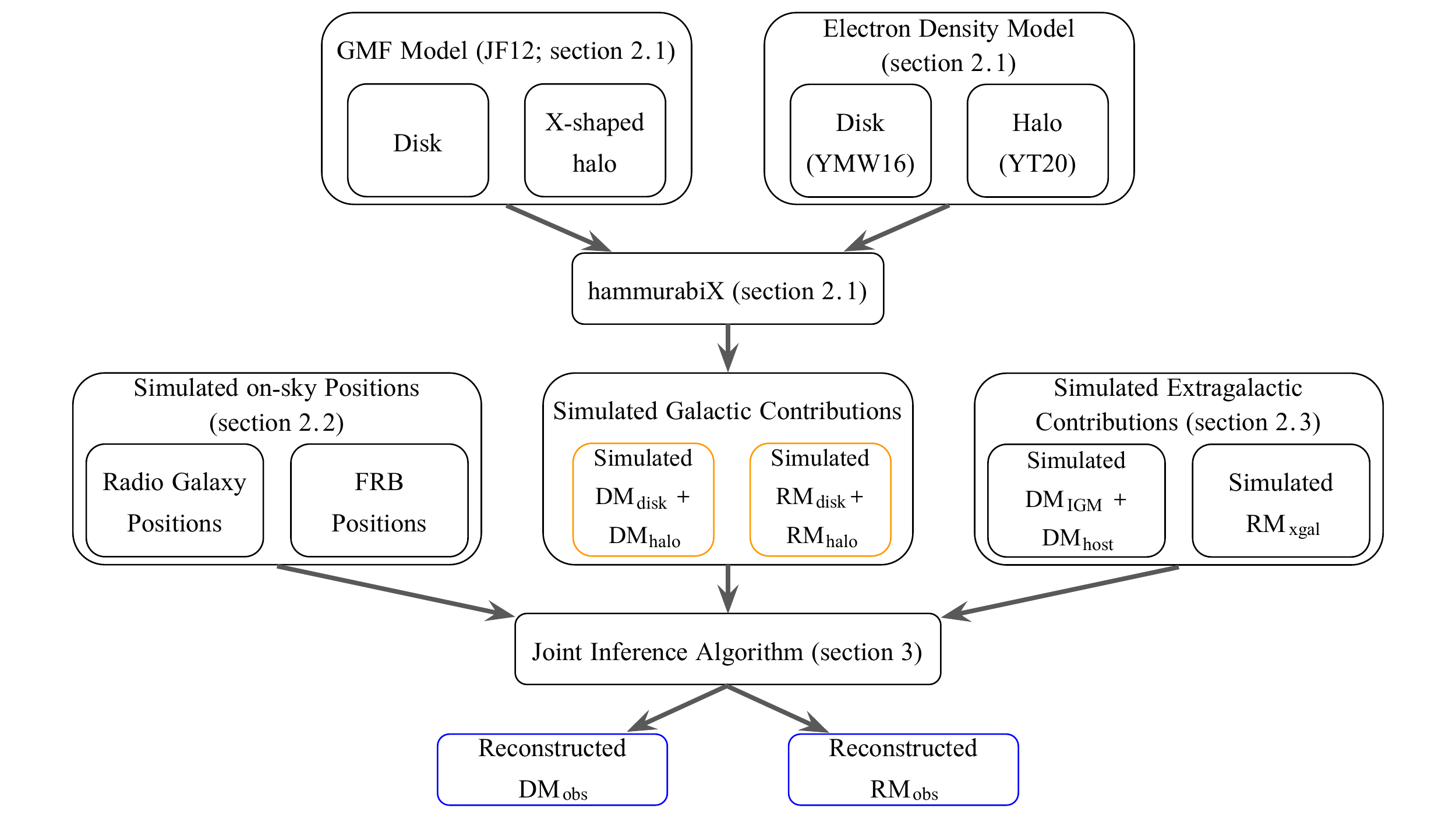}
    \caption{A flowchart summarizing the assumed models, methods, and outputs of our simulated data generation process (section \ref{simobs}) and the joint inference algorithm (section \ref{inference}). DM and RM components follow the same breakdown as equations \ref{eq:dm2} and \ref{eq:rm2}. We highlight the Galactic maps we are attempting to recreate in orange and the reconstruction results in blue. Note that $\left<B_{\parallel}\right>$ is not a direct output of {\tt hammurabiX} or the joint inference algorithm and is instead computed from the RM and DM by using equation \ref{eq:B}.}
    \label{fig:flowchart}
\end{figure*}

This section details the various steps taken to create our simulated DM and RM observations. These steps are summarized as a flowchart in Figure \ref{fig:flowchart}.

\subsection{HammurabiX \& Input Models}
The {\tt hammurabiX}\footnote{\href{https://github.com/hammurabi-dev/hammurabiX}{https://github.com/hammurabi-dev/hammurabiX}} (V2.4.1) package \citep{2020ApJS..247...18W} is an open source astrophysical magnetic field simulator that incorporates user input 3D models of different components of the MW disk ISM and halo, such as magnetic fields, and thermal and relativistic electrons. Given input models for these components, the code utilizes the {\tt HEALPix} library to execute efficient LOS integrals through the simulated 3D Galactic model by using multi-layered spherical shells with adaptable resolutions. {\tt HammurabiX} integrates along the LOS to compute the synchrotron emission component of Stokes $I$, $Q$, and $U$, and the Faraday depth and DM are then calculated from discrete forms of equations \ref{eq:RM} and \ref{eq:DM}. For the purposes of this work, we utilize {\tt hammurabiX} to generate simulated maps of RM and DM by inputting pre-existing models for the GMF and thermal electron density. Below, we discuss the specifics of the assumed models in this study; this collection of models was chosen because of their flexible implementation in {\tt hammurabiX}, which allows for alterations to each model parameter.

For the GMF model, we adopt the formulation laid out by \cite{2012ApJ...757...14J} (hereafter JF12). The JF12 model is comprised of: (i) a disk component between Galactic radii $r$ of 3 kpc and 20 kpc, containing a purely azimuthal field at 3 kpc < $r$ < 5 kpc and eight logarithmic spiral regions beyond 5 kpc; (ii) a toroidal halo component with an exponential scale height and independent amplitudes parameters for the north and south sections of the field; (iii) an X-shaped poloidal halo component$-$which is not included in many other GMF models$-$replicating similar behaviour seen in radio observations of edge-on galaxies \citep[e.g.][]{2009RMxAC..36...25K, 2009Ap&SS.320...77B, 2020A&A...639A.112K}. For a set of best-fit model parameters see Table 1 of \cite{2012ApJ...757...14J}.

The JF12 model was originally written in {\tt hammurabi} \citep{Waelkens2009}, a predecessor to {\tt hammurabiX}. The model was made available in the V2.4.1 release of {\tt hammurabiX}.  The coherent magnetic field is consistent in both implementations, although the {\tt hammurabiX} implementation creates generally smoother maps due to changes in the numerical integration routines.

The thermal electron density model utilized in this work incorporates separate models for the disk component \citep{2017ApJ...835...29Y} (hereafter YMW16) and halo component \citep{2020ApJ...888..105Y} (hereafter YT20). YMW16 is made up of: (i) a thick disk representing the diffuse ionized medium; (ii) a thin disk which models the region of increased gas density and star formation a few kpc out from the Galactic center, dubbed the "molecular ring"; (iii) the Galactic center; (iv) a set of spiral arms \citep[for specific spiral-arm parameters see Table 1 of][]{2017ApJ...835...29Y}; (v) various local features (e.g. the Gum Nebula or Local Bubble). For this work, we elect to remove the Gum Nebula from the model as its implementation within {\tt hammurabiX} (V2.4.1) creates unintended sharp features in the on-sky projection of key observables, such as RM and DM, that are unphysical. We note that the choice to remove the Gum Nebula from the model has no impact on this study and its results. In regards to the halo model, YT20 is a sum of two exponential components, a compact disk-like component and an extended spherical halo.

\subsection{Simulating Radio Galaxy \& FRB Populations}
\begin{figure}
    \centering
    \includegraphics[scale=0.195]{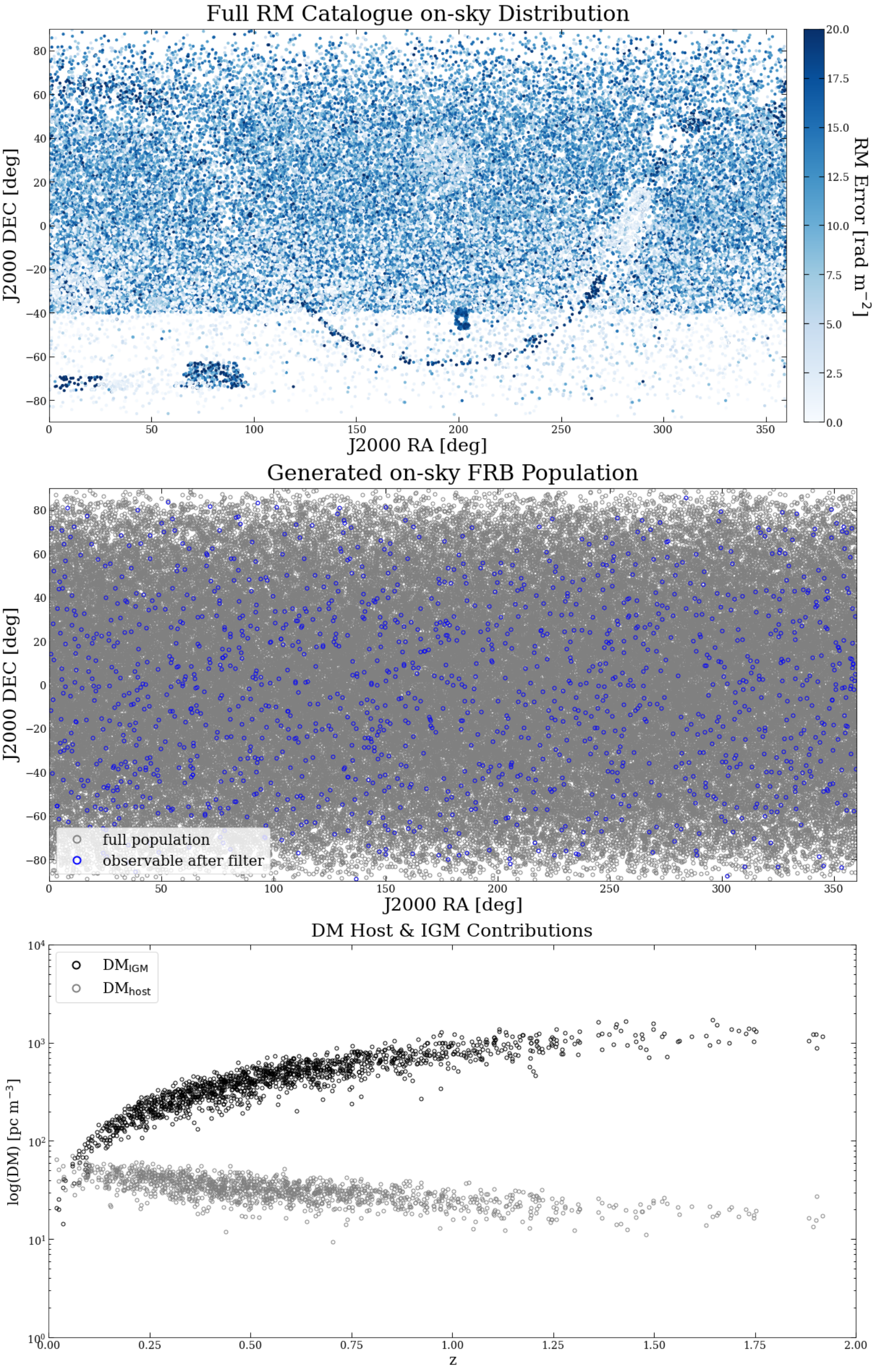}
    \caption{A projection of the assumed radio galaxy and FRB populations in J2000 celestial coordinates and a breakdown of the assumed extragalactic DM model. (Top) The on-sky distribution of the assumed radio galaxy population in this work. In total there are 50101 extragalactic point sources that are colored as a function of the corresponding error in RM measurements. (Middle) The on-sky distribution of the generated FRB population spanning redshifts 0 < $z$ < 2. Grey points represent the total generated populations (100000 sources) and the blue points are the sub-sample that pass our filter (1308 sources). (Bottom) Contributions to the total observed DM along a given LOS from extragalactic sources as a function of redshift. Contribution from host galaxies are plotted as grey points while black points correspond to the IGM contribution.}
    \label{fig:agnfrbpop}
\end{figure}

To simulate LOS RM and DM observations, we require a method of generating typical radio galaxy and FRB populations, respectively. To obtain a population of radio galaxies in line with current understanding and observations, we take advantage of the aforementioned catalogue of extragalactic point sources (\textcolor{blue}{Van Eck et al.} \textcolor{blue}{in prep}; V0.2.1). However, the simulated LOS observables will be determined by the assumed field models and thus we only extract the on-sky positions, in both celestial and Galactic coordinates systems, along with measured RM errors. The measured RM itself is not taken from the pre-existing catalogue and will be evaluated in a later step based on equation \ref{eq:RM} and assumed GMF and $n_e$ models. Moreover, any known pulsars (35 out of 50207 entries) and data points without a reported RM error (71 out of 50207 entries) are removed from the sample. The remaining 50101 point sources comprise our radio galaxy population and are plotted with respect to their measured RA, DEC, and RM$_{\mathrm{err}}$ in Figure \ref{fig:agnfrbpop}\textcolor{blue}.

The {\tt frbpoppy}\footnote{\href{https://github.com/davidgardenier/frbpoppy}{https://github.com/davidgardenier/frbpoppy}} (V2.1.0) package is a tool for conducting FRB population synthesis \citep{2019A&A...632A.125G}. Our goal with this package is to generate a simulated sample of FRBs that would be realistically observable by current radio instruments. Typically, assuming a 1~ms burst duration, observed FRBs have luminosities on the order of $\sim 10^{8} - 10^{9}$ L$_{\odot}$ (or $\sim 10^{41} - 10^{42}$ erg s$^{-1}$) \citep[see Figure 2 of][]{2020Natur.587...54T} and are inferred to be occuring in galaxies with a cosmological distance scale, $z \sim 1$ \citep{2020ApJ...888..105Y}. Therefore, we utilize {\tt frbpoppy} to generate three independent sets of populations with roughly 1000, 10000, and 50000 observable FRBs each. The varying sized datasets enable us to analyze the ability of our joint inference technique to produce accurate results with increasingly larger samples of FRBs. Likewise, the range of population sizes encompasses the approximate number of FRB observations available within the next few years and up to the stage where they are commensurate with the current RM catalogue. The population is distributed over a conservative range of redshifts 0 < $z$ < 2 and luminosities 10$^{40}$ erg s$^{-1}$ < $L$ < 10$^{45}$ erg s$^{-1}$ bracketing the typical quantities from \cite{2020ApJ...888..105Y} and \cite{2020Natur.587...54T}, respectively. FRB luminosities are drawn from a uniform distribution spanning this range. We follow the default FRB spectral index and pulse width distributions provided in {\tt frbpoppy}. The spectral index $\gamma$ follows a Gaussian distribution with mean $\bar{\gamma}$ = $-$1.4 and standard deviation $\sigma_{\gamma}$ = 1 \citep[][and references therein]{2019A&A...632A.125G}. Meanwhile, the FRB pulse width $w$ follows a uniform distribution spanning 0 ms < $w$ < 10 ms. It should be noted that there are a number of effects that modify the pulse width before it is measured on Earth; for a detailed discussion of these effects see Section 3 of \cite{2019A&A...632A.125G}. Additionally, we assume a uniform distribution of sources on the sky and a constant number density of sources per comoving volume.

Next, we apply a filtering criterion to the FRB population with respect to a selected instrument or survey to draw out a sub-sample of FRBs that would be realistically observable. Since the CHIME/FRB project has, so far, collated the largest set of FRB data, we elect to use a filter that resembles CHIME/FRB specifications. While the exact survey design of CHIME is complicated and difficult to replicate, a simple version of its system parameters is implemented into {\tt frbpoppy}. Here, we present a list of some relevant parameters (for a full list of parameters, see \href{https://github.com/davidgardenier/frbpoppy}{https://github.com/davidgardenier/frbpoppy}): frequency range: 400$-$800~MHz, receiver temperature: 50~K, sampling time: 1~ms, integration time: 360~s, and detection threshold signal-to-noise ratio: 10. In reality, the sky distribution of CHIME/FRB observations ranges between approximately $-$11~deg < DEC < 90~deg and is not uniformly distributed due to variations in source transit duration as a function of declination \citep{2018ApJ...863...48C, 2021ApJS..257...59A}. Additionally, the sensitivity of the observations also varies with declination. However, in this work we do not attempt to replicate the sky distribution of CHIME/FRB sources but rather generate a set of full sky simulated observations that is representative of typical FRB observations. Consequently, we apply this filter over the full range of RA and DEC and maintain a uniform sky distribution of simulated sources. An example of a generated FRB population, with a total of 100000 sources, is plotted in Figure \ref{fig:agnfrbpop}. In this example, approximately 1.3\% of the total population successfully pass the filtering criteria and are overplotted as blue markers.

\subsection{Simulating LOS Observations}
With the radio galaxy and FRB populations in place, we must now account for the DM and RM contributions from Galactic and extragalactic components of the foreground sight-lines. For most extragalactic RM sources, the net polarization angle rotation is dominated by effects originating from the MW's ISM \citep[e.g.][]{2010MNRAS.409L..99S, 2015A&A...575A.118O}. \cite{2010MNRAS.409L..99S} modelled the width of the RM distribution, $\sigma_{\rm{RM}}$, for NVSS RM catalogue sources as a dominant Galactic contribution ($\bar{\sigma}_{\rm{RM,MW}} \sim 8$ rad m$^{-2}$) that is amplified at smaller Galactic latitudes as $1/\rm{sin}|b|$ and a small extragalactic component ($\bar{\sigma}_{\rm{RM,EG}} \sim 6$ rad m$^{-2}$) independent of Galactic latitude: $\sigma_{\rm{RM}}(b)^2 = \left( \frac{\bar{\sigma}_{\rm{RM,MW}}}{\rm{sin}|b|} \right)^2 + \bar{\sigma}_{\rm{RM,EG}}^2$. Furthermore, \cite{2015A&A...575A.118O} estimate the contribution from non-Galactic media using a statistical noise model \citep[for a detailed discussion of the noise model see section 2.3.2 of][]{2020A&A...633A.150H}. The fundamental idea of this process is to use on-sky correlations as the basis for discriminators between Galactic and extragalactic contribution. Based on this formulation, \cite{2015A&A...575A.118O} place an upper limit on the standard deviation of extragalactic Faraday depth contribution at $\sigma_{\rm{RM,EG}} \sim 7$~rad m$^{-2}$. Their result is in close agreement with \cite{2010MNRAS.409L..99S} and shows that the Galactic component is largely dominant in RM observations. Conceptually, this stems from the $B_{\parallel}$ dependence seen in equation \ref{eq:RM} which is not always pointed in the same direction along the entire path length and can often reverse along the LOS when the path length is much larger than the magnetic field scales (e.g. in the case of extragalactic sources). This is not the case for extragalactic DM contributions because the integral of $n_e$ is a monotonically increasing function of distance.

In this work, the Galactic RM contribution along a given LOS is computed as a simulated observable from {\tt hammurabiX}, assuming the combination of previously discussed models (YMW16, YT20, and JF12). We independently incorporate an extragalactic component by sampling from a normal distribution centered at zero with a standard deviation of $\sigma_{\mathrm{RM,EG}} = 7$~rad m$^{-2}$ for each simulated LOS observation.

Similarly, the Galactic ISM and halo components of DM are output from {\tt hammurabiX}, derived from the input $n_e$ models (YMW16 and YT20). However, in the case of DM, there is significant contributions from the host galaxy, $\rm{DM}_{\rm{host}}$, and IGM, $\rm{DM}_{\rm{IGM}}$. In this regard, we follow the procedure described by \cite{2020Natur.581..391M}, who find that $\rm{DM}_{\rm{IGM}}$ is expected to have a strong redshift dependence, while the value of $\rm{DM}_{\rm{host}}$ as observed from Earth is weighted by ($1 + z_{\rm{FRB}}$)$^{-1}$. In particular, these quantities are approximated as:
\begin{equation}
\rm{DM}_{\rm{host}} = 50 (1 + z_{\rm{FRB}})^{-1}~\mathrm{pc}~\mathrm{cm}^{-3}\label{eq:dmhost} \,,
\end{equation}
\begin{equation}
\left<\rm{DM}_{\rm{IGM}}\right> = \int_{0}^{z_{\rm{FRB}}} \frac{c \widetilde{n}_e(z) dz}{H_0(1 + z)^2\sqrt{\Omega_{m}(1 + z)^3 + \Omega_{\Lambda}}}
\label{eq:dmigm} \,,
\end{equation}
where $\widetilde{n}_e = f_d \rho_b(z) m_p^{-1}(1 - Y_{\rm{He}}/2)$, with proton mass $m_p$, helium mass fraction $Y_{\rm{He}}$ that is assumed to be doubly ionized, the fraction of cosmic baryons in diffuse ionized gas $f_d(z)$, $\Omega_{m}$ and $\Omega_{\Lambda}$ are the current matter and dark energy densities of the Universe, respectively, and $\rho_b(z) = \Omega_{b} \rho_{c,0} (1 + z)^3$ with $\rho_{c,0} = 3 H_0^2/8 \pi G$. For the subset of arcsecond and sub-arcsecond localized FRBs, this leads to an empirical relationship between $\rm{DM}_{\rm{IGM}}$ and $z$, dubbed the ``Macquart relation'' \citep[see Figure 2 from][]{2020Natur.581..391M}. In their analysis, the estimated DM contribution from the MW and the FRB host galaxy has been subtracted from $\rm{DM}_{\rm{obs}}$. The localized FRB sample is in agreement with the expected $\rm{DM}_{\rm{IGM}}-z$ relation (see equation \ref{eq:dmigm}) for \textit{Planck} 15 cosmology (i.e. $\Omega_b = 0.0486$ and $H_0 = 67.74$ km s$^{-1}$ Mpc$^{-1}$) \citep{2016A&A...594A..13P} after accounting for the scatter from the IGM.

In this work, we adopt the formalism described above to determine the extragalactic contribution to DM$_{\rm{obs}}$. Namely, DM$_{\rm{host}}$ follows a Gaussian model with the mean being equivalent to equation \ref{eq:dmhost} and a standard deviation of $\sigma_{\rm{DM,host}} = 10 (1 + z_{\rm{FRB}})^{-1}$~pc cm$^{-3}$. Similarly, we model our IGM contribution to follow the linear $\rm{DM}_{\rm{IGM}}-z$ relation derived from equation \ref{eq:dmigm} with a 20\% spread as a function of redshift (i.e. $\sigma_{\rm{DM,IGM}} = 0.2 \left<\rm{DM}_{\rm{IGM}}\right>(z)$~pc cm$^{-3}$). For the same example population of observable FRBs shown in the second panel of Figure \ref{fig:agnfrbpop}, we plot the distribution of simulated DM$_{\rm{host}}$ and DM$_{\rm{IGM}}$ contributions as a function of redshift in the final panel of Figure \ref{fig:agnfrbpop}. There are examples in literature of studies that examine additional sources of extragalactic DM contributions. For example, \cite{2019MNRAS.485..648P} study DM contributions from the highly ionized Local Group Medium (LGM) and generate sky-projection maps of the DM$_{\rm{LGM}}$ and DM contribution from M31 and the Magellanic Clouds. However, in this work we do not include these components and rather consider a simple model in which the extragalactic DM is built up entirely from contributions from the host galaxy and IGM.

\section{Joint Inference Reconstruction} \label{inference}
In this section, we describe the joint inference setup to reconstruct the Galactic RM and DM skies from extragalactic point source measurements. Depending on their location and density on the sky, these data sets probe the Galactic sky on various scales and environments. Our a priori knowledge, especially on the Galactic DM sky, is rather limited and usually constrained to the largest scales \citep[e.g. see][]{2002astro.ph..7156C, 2017ApJ...835...29Y} or along specific LOS constrained by pulsars. This means we have no reliable parametric model at hand that we can expect to be an accurate representation of the underlying field on all scales and locations probed by our data sets. Hence, we opt to infer both skies non-parametrically by utilizing generic models with a large number of degrees of freedom that are only constrained by the data and generic assumptions on smoothness and the expected range of values. This is done by formulating the problem within IFT, a Bayesian inference framework useful for high dimensional problems \citep{2019AnP...53100127E}. The resulting posterior distribution, which, depending on the resolution of the sky maps, can have millions of degrees of freedom, is evaluated using a variational Bayesian optimization scheme called Metric Gaussian Variational Inference \citep{MGVI}.

In the next paragraphs, we summarize the components of our model that are specific to this work. We start our description by connecting the data $d_s$ (with $s \in (\rm{RM,DM})$) to the respective Galactic sky maps $s_{\rm{gal}}$, which are defined on a {\tt HEALPix} \citep{healpix} grid on the unit sphere,
\begin{equation}
\label{eq:data_new}
d_s = \mathcal{R}s_\mathrm{gal} +  \mathcal{R}s_\mathrm{xgal} + n_s, 
\end{equation}
where $\mathcal{R}$ is a projection operator on the data vector, $s_\mathrm{xgal}$ describes the extragalactic component, and $n_s$ is the observational noise. The latter is assumed to be drawn from a zero mean Gaussian with known diagonal covariance $\sigma^2_{d_s}$. At this point, we can identify three issues, which need to be addressed in our modeling:
\begin{enumerate}
\item While we are interested in $s_\mathrm{gal}$, equation \ref{eq:data_new} is usually not invertible even in the noiseless case (i.e. not every pixel of the sky maps will be constrained by data). Thus, we need to infer an interpolation kernel that is able to constrain pixels by exploiting correlations of nearby data points; 
\item The extragalactic contribution, particularly in DM, is often as strong as or stronger than the Galactic component. Therefore, we require a method of constraining and separating the extragalactic contributions to better estimate the Galactic structure;
\item We need a physically plausible inference model for both $\mathrm{RM}_{\rm{gal}}$ and $\mathrm{DM}_{\rm{gal}}$, which is capable of representing known large scale features, such as the Galactic disk, but is also flexible enough to characterize unknown structures hidden in the data. The model should furthermore be able to efficiently exploit the correlations between the RM and DM maps.
\end{enumerate}
Issue (i) touches on an integral feature of IFT and the Numerical Information Field Theory Python package (NIFTy) \citep{asclnifty5}, namely the modeling of correlations. We parameterize each sky map as a non-linear combination of Gaussian sky maps, which have a parametrizable correlation structure in terms of power spectra. For the details of the power spectrum modeling we refer the reader to \cite{arras}.

In regard to issue (ii), we refrain from explicitly modelling the extragalactic component (for both RM and DM) in the inference but instead take several steps to minimize the effects of extragalactic contribution on the reconstructed results by means of data reduction and the derivation of a correction factor. For a detailed discussion of these steps, see Sections \ref{results} and \ref{discussion}, respectively.

\begin{figure}
\centering
\begin{tikzpicture}[auto]
  \matrix[ampersand replacement=\&, row sep=0.5cm, column sep=0.5cm] {
    \&
    \&
    \node [block] (rho) {$\rho$};
    \&
    \&
	\node [block] (chi) {$\chi$};    
    \&
    \&
    \\
    \node [block] (n_dm) {$n_\mathrm{DM}$};
    \&
    \&
    \node [block] (dm) {$\mathrm{DM}$};
    \&
    \&
    \node [block] (rm) {$\mathrm{RM}$};
    \&
    \&
    \node [block] (n_rm) {$n_\mathrm{RM}$};
    \\
    \&
    \&
    \node [block] (d_dm) {$d_\mathrm{DM}$};
    \&
    \&
    \node [block] (d_rm) {$d_\mathrm{RM}$};
    \&
    \&
    \&
    \\ 
    }; 
    \path [line] (chi) -- node{}(rm);
    \path [line] (rho) -- node{}(rm);
    \path [line] (rho) -- node{}(dm);
    \path [line] (rm) -- node{}(d_rm);
    \path [line] (n_rm) -- node{}(d_rm);
    \path [line] (dm) -- node{}(d_dm);
    \path [line] (n_dm) -- node{}(d_dm);

\end{tikzpicture}
\caption{\label{fig:tikz} A simplified model tree for the joint RM/DM inference scheme. In reality, there exist several layers above the $\rho$ and $\chi$ nodes pertaining to the implicit correlation structure modeling.}
\end{figure}
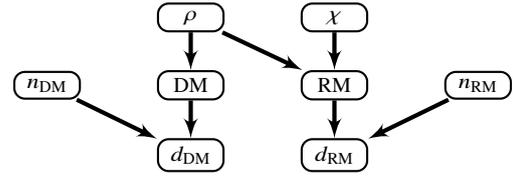

Finally, regarding issue (iii), the modeling of the Galactic RM and DM skies needs to be discussed. For the RM sky, we follow the recipe of \cite{2020A&A...633A.150H} and \cite{h21} and model the map as a combination of two components,
\begin{equation}
\mathrm{RM_{gal}} = 0.812 e^{\rho}\chi. \label{eq:rm_inference_model}
\end{equation}
Both $\rho$ and $\chi$ are Gaussian sky maps with a priori unknown power spectra. This is mostly motivated by the need to model both strong variations in RM amplitude (mostly via the log-normal $\rho$ map), and the sign of the RM sky (via the $\chi$ map). This models ties back in to equation \ref{eq:B} insofar as the thermal electron density and magnetic field vector are uncorrelated. 
In this case, $e^\rho$ might be viewed as a proxy for the morphology of the DM sky, which dominates the sky structure of the RM amplitude \citep{2020A&A...633A.150H}. 
Similarly, the $\chi$ map can then be seen as a proxy for the averaged magnetic field sky, as amplitude structures correlated with the sign are most likely also represented in this map and under the assumption that these are then most likely caused by the magnetic field. 
Unfortunately, no further conclusions on absolute scales or the reliability of specific structures can be drawn due to degeneracy between the two maps, if the model is constrained using RM data only as in \cite{h21}. 
However, this changes if we add the DM data to constrain the amplitude field by assuming, 
\begin{equation}
\mathrm{DM_{gal}} = e^{\rho} \,, \label{eq:dm_inference_model}
\end{equation}
as a model for the DM sky, which is also proposed by \textcolor{blue}{Hutschenreuter} (\textcolor{blue}{in prep}). As the DM map is also parametrized by $\rho$, both sky maps are now connected in the inference and any change in either map will immediately be reflected in the other. From this it follows that, if we assume no correlations between $n_e$ and $\mathrm{B}_\parallel$, the posterior $\chi$ map should represent our best estimate on the averaged LOS component of the GMF. The corresponding hierarchical tree of this joint inference model is depicted in Figure \ref{fig:tikz}. The inference algorithm is implemented in NIFTy version 7. 

\section{Results} \label{results}
The results of the joint inference reconstruction for various input models and testing are presented throughout this section. All on-sky maps are presented in the form of {\tt HEALPix} Mollweide projections with {\tt nside} = 128 (i.e. with a total of 196608 pixels or, equivalently, a resolution of $\sim$ 27.5 arcminutes) and in Galactic coordinates. The reconstructed maps are directly compared to the underlying fields by way of contrasting against the all-sky {\tt hammurabiX} outputs for DM, RM, and $\left<B_{\parallel}\right>$. This method of analysis tests not only the accuracy of the reconstruction but also how well we are able to disentangle the Galactic and extragalactic contributions from the simulated observations. The key results are included in this section, but complete reconstruction information, including corresponding reconstruction errors, are located in Appendix \ref{recon_results}.

\subsection{Baseline Galactic Model}
\begin{figure*}
    \centering
    \includegraphics[scale=0.235]{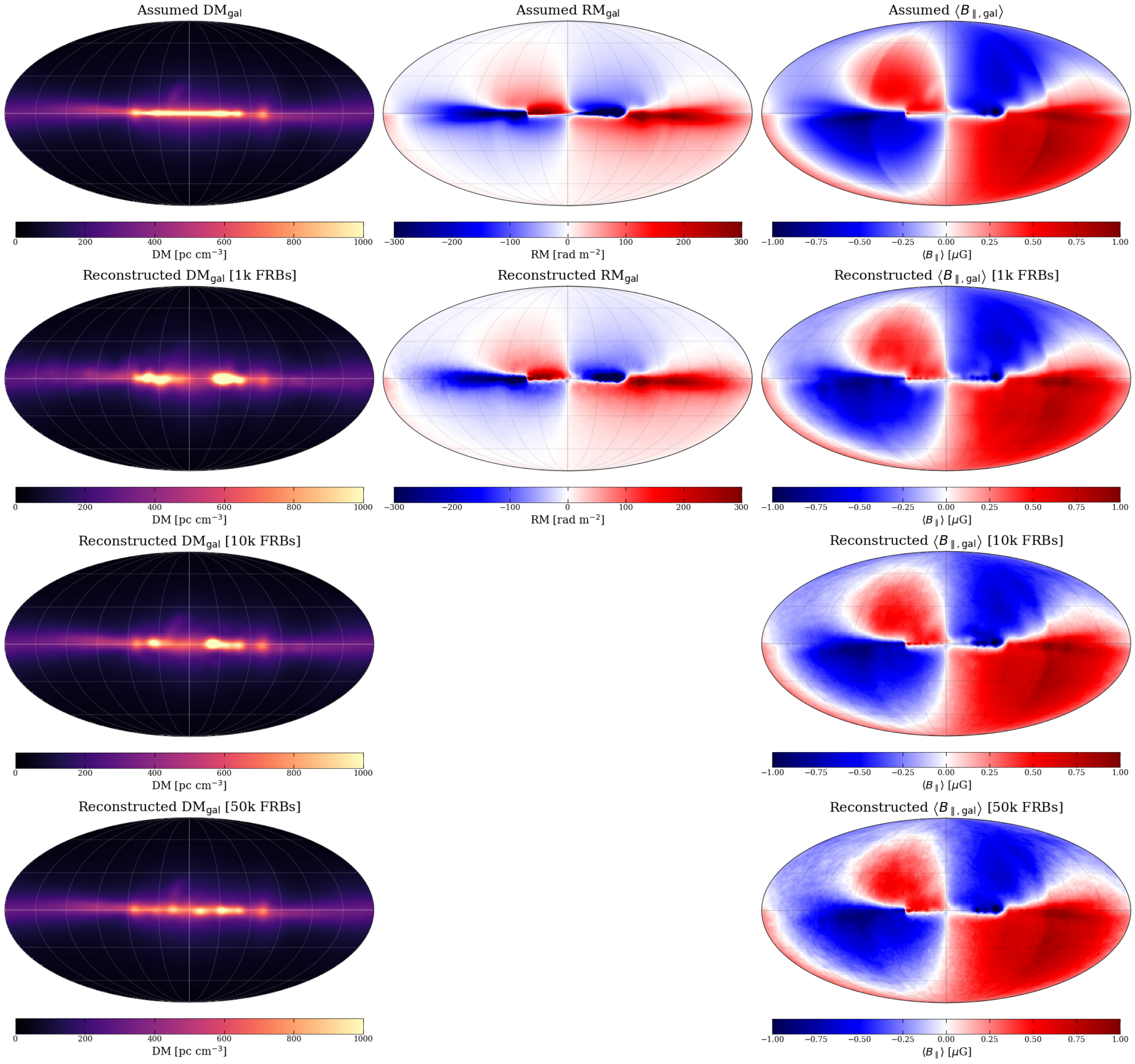}
    \caption{Set of reconstructed DM, RM, and $\left<B_{\parallel}\right>$ maps in comparison to the expected on-sky observables from {\tt hammurabiX}. All maps are plotted as a Mollweide projection in Galactic coordinates with a resolution of $\sim$ 27.5 arcminutes. The underlying models used for reconstruction explicitly have no extragalactic contributions. The expected total Galactic contribution for each observable is shown in the top row and subsequent rows depict the results of the reconstruction algorithm on progressively larger sets of FRB data (1000, 10000, and 50000). The radio galaxy population has a fixed size and therefore only the RM reconstruction for one run of the joint inference algorithm is presented.}
    \label{fig:maps_gal}
\end{figure*}

To test the capability of the reconstruction algorithm, independently of how well we can constrain the extragalactic component of the observables, we begin by applying the algorithm to a set of data that contains only the Galactic information along each LOS. The extragalactic component is explicitly set to zero at each LOS; while this is unphysical, it will provide a useful baseline for the best possible reconstruction results. Each column of Figure \ref{fig:maps_gal} contains the on-sky DM, RM, and $\left<B_{\parallel}\right>$ (computed according to equations \ref{eq:RM}, \ref{eq:DM}, and \ref{eq:B}) information, respectively. The first row shows the {\tt hammurabiX} outputs based on our assumed Galactic models and each consequent row presents the reconstructed map of each observable for varying sized data sets. As was mentioned in Section \ref{simobs}, unlike the FRB population, the radio galaxy population that we have adapted in this study has a fixed size and therefore we only present the RM reconstruction for one run of the joint inference algorithm. The changes in the RM reconstruction are negligible as we vary the size of the FRB population between runs. From Figure \ref{fig:maps_gal} alone, it can be difficult to see the differences between the reconstruction results and the corresponding {\tt hammurabiX} output. To quantitatively measure these differences, we analyze the accuracy of the reconstruction results below. In addition, we generate plots of the DM maps in a log base 10 scale (to better visualize low-DM regions away from the Galactic plane) and provide difference maps for the RM and $\left<B_{\parallel}\right>$ reconstructions in Appendix \ref{recon_results}.

With a population of radio galaxies distributed identically to that of the observed sample, we are able to reconstruct the RM-sky to an extremely high degree of precision. The DM reconstruction for 50000 FRBs is also largely in good agreement with the expected DM-sky. Especially with regard to characterizing the shape of the large scale structure in the Galactic plane (for example the high DM closer to the Galactic centre (GC) followed by lower DMs away from the GC) the algorithm performs exceptionally. In the outer regions of the disk (60 deg $\lesssim \ell \lesssim$ 300 deg), the absolute difference between the reconstructed DM and the {\tt hammurabiX} map is $\lesssim$ 20~pc cm$^{-3}$. The largest discrepancy is in approximating the magnitude of the DM in the Galactic plane near the GC. Here, the assumed $n_e$ model has DMs $\gtrsim$ 10$^{3}$~pc cm$^{-3}$ uniformly in the plane but the joint inference algorithm appears to identify high DM simulated observations in this region as localized structure. Therefore, the reconstructed map underestimates the DM by $\sim$ 10$^{2}$ $-$ 10$^{3}$~pc cm$^{-3}$ in small patches of the inner Galactic plane ($|b| \lesssim$ 5 deg, $\ell \lesssim$ 60 deg and $\ell \gtrsim$ 300 deg). Local features, such as spiral arms manifesting as roughly circular over-densities in the disk, are also easily identifiable in the reconstruction. Moving off the Galactic plane, the reconstruction agrees with the model to within $\lesssim$ 5~pc cm$^{-3}$ (for 5 deg $ \leq |b| \leq$ 30 deg) and to within $\lesssim$ 1~pc cm$^{-3}$ (for $|b|$ > 30 deg). 

As we move to smaller data sets (10000 and 1000 FRBs, respectively), the results maintain a lot of the large scale structure but the accuracy of the reconstruction begins to decline. In particular, with only 1000 data points, the coverage becomes patchy at points throughout the Galactic plane and the local features (spiral arms and the Loop I region) are difficult to recognize. Even in the outer regions of the Galactic plane, which were well estimated with the 50000 FRB sample, we begin to see DMs that differ from the {\tt hammurabiX} output by $\sim$ 10$^{2}$~pc cm$^{-3}$. Off the plane, the 1000 FRB reconstruction differs from the {\tt hammurabiX} output by $\lesssim$ 12~pc cm$^{-3}$ (for 5 deg $ \leq |b| \leq$ 30 deg) and by $\lesssim$ 2~pc cm$^{-3}$ (for $|b|$ > 30 deg).

Following equation \ref{eq:B}, the accuracy of the reconstructed $\left<B_{\parallel}\right>$-sky is directly dependent on how well we are able to characterize the DM and RM, respectively. Since the only significant discrepancies from the expected values are in the DM reconstruction, the accuracy of the $\left<B_{\parallel}\right>$ maps are reflective of the accuracy of the corresponding DM-sky and we observe the same trends with the 10000 and 1000 FRB reconstruction. For the 50000 FRB reconstruction, there were some small regions in the inner Galactic plane that were within $\lesssim$ 0.3~$\mu$G of the corresponding {\tt hammurabiX} map, but most of the Galactic plane and halo was within $\lesssim$ 0.1~$\mu$G. At 1000 FRBs, the patchy coverage is again present (similar to the corresponding DM map) and we find discrepancies of $\lesssim$ 0.3~$\mu$G from the expected value throughout most of the disk and $\lesssim$ 0.15~$\mu$G in the halo.

\subsection{Full Model Reconstruction}
\begin{figure*}
    \centering
    \includegraphics[scale=0.175]{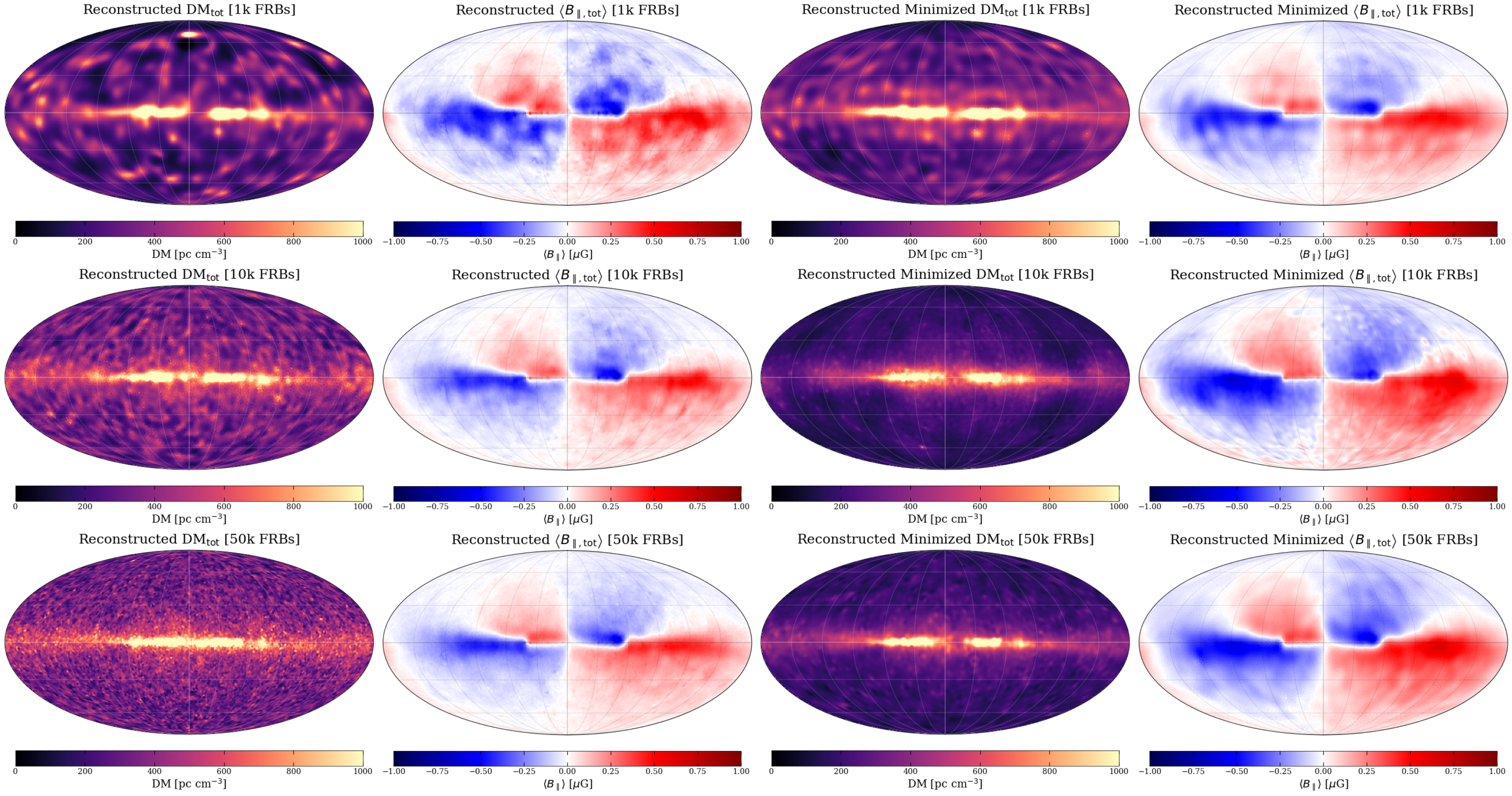}
    \caption{As for the first two columns of Figure \ref{fig:maps_gal} but now considering the full extent of DM contributions from extragalactic sources as well. The reconstructed RM is the same as Figure \ref{fig:maps_gal} and is therefore not plotted here. Again, each subsequent row represents the reconstruction results from successively larger sized data sets. The first two columns are results from the raw (1000, 10000, 50000 FRB) data sets and the last two columns are after they have been filtered to minimize the extragalactic DM contributions.}
    \label{fig:maps_standard}
\end{figure*}

Administering the joint inference algorithm on the full simulated LOS observations (now including the extragalactic contributions) drastically changes the results from the previous section. Dissimilar from what we see in RM, the extragalactic DM contribution is significant when compared to Galactic contributions from the MW's ISM and halo. The first two columns of Figure \ref{fig:maps_standard} show the reconstructed DM and $\left<B_{\parallel}\right>$ maps; similar to Figure \ref{fig:maps_gal}, successive rows illustrate the results of the algorithm when applied to increasingly smaller FRB data sets. The reconstruction for 50000 FRBs shows a lot of the large scale Galactic DM structure visible in the corresponding panel of Figure \ref{fig:maps_gal} and local DM structure is still, to a lesser degree, discernible. However, the extragalactic components add a mean DM contribution of approximately 300~pc cm$^{-3}$ to the reconstruction results that does not show any preferred structure as a function of Galactic coordinates and is generally isotropic. Thus, the sharp local features noted in the previous section are drowned out to an extent. For the smallest data set (1000 FRBs), where the algorithm must interpolate over larger areas of the sky that lack FRB data, the extragalactic contribution begins to appear weakly structured. However, we know that this extragalactic contribution should not be correlated over Galactic coordinates and it is therefore a sign that there are an insufficient number of data points for an accurate reconstruction, especially away from the Galactic plane. The overall effect of overestimating the Galactic DM due to extragalactic contributions translates to an underestimation of the parallel magnetic field component $\left<B_{\parallel}\right>$ (following equation \ref{eq:B}), particularly at higher latitudes.

To reduce the aforementioned effects and recover more of the Galactic DM and $\left<B_{\parallel}\right>$ structure, we employ a technique to minimize the extragalactic DM in equal-area portions of the sky. To accomplish this, we utilize a low resolution Mollweide projection {\tt HEALPix} grid, which divides the sky into equal-area sub-sections. For the largest simulated data set, we use a grid with {\tt nside} = 16 (i.e. 3072 sub-sections with a mean spacing of 3.6645 deg) and from all LOS observations encompassed in an individual sub-section, we only keep the single LOS with the minimum DM$_{\rm{tot}}$ (likely the nearest FRB with the lowest observed extragalactic DM contribution). Repeating this over the whole sky, our filtered FRB population then contains a total of 3072 data points that are roughly evenly distributed across the sky. The same process is used to filter the smaller populations, although the {\tt HEALPix} grid resolution is decreased to {\tt nside} = 8 (i.e. 768 sub-sections with a mean spacing of 7.3290 deg) to compensate for the smaller data sets, and the filtered populations contain 768 (initially 10000) and 563 (initially 1000) FRBs, respectively. From the last two columns of Figure \ref{fig:maps_standard}, we see that this step does improve both DM and $\left<B_{\parallel}\right>$ reconstructed maps and we better recover some of the local features of the Galactic structure. Specifically, the mean DM contribution from extragalactic sources is decreased by $\sim$ 150~pc cm$^{-3}$ and $\left|\left<B_{\parallel}\right>\right|$ is increased by roughly between $\sim$ 0.1 to 0.3~$\mu$G over most of the sky. For a sample of FRBs that are uniformly distributed over the entire sky, as we increase the number of FRBs in the sample, we increase the number of simulated LOS observations per equal-area sub-section of the sky and thus increase the probability of each sub-section containing a nearby, low DM FRB. Therefore, the filter becomes more effective as we increase the number of FRBs in our data set. Hence, this filter is most effective on the sample of 50000 FRBs and becomes progressively less effective for the smaller data sets. For the smallest set (1000 FRBs) we are only able to discard roughly half of the high DM observations without significantly impacting the resolution of the reconstruction. It should also be noted that this technique discards information pertaining to very small scale Galactic DM structure (i.e. Galactic DM structure with a smaller angular resolution than the equal-area sky patches). In this work, we are focused on large scale structure and this effect does not meaningfully influence our results.

However, this method of constraining the extragalactic DM is not perfect and even after this step, the reconstructed DM ($\left<B_{\parallel}\right>$) map is still overestimated (underestimated) when contrasted with the expected Galactic contributions from the first row of Figure \ref{fig:maps_gal}. Namely, there is still a mean extragalactic DM contribution of $\sim$ 150~pc cm$^{-3}$, which we anticipate from the bottom panel of Figure \ref{fig:agnfrbpop}. In Section \ref{discussion}, we attempt to correct for the remaining extragalactic contribution by deriving a possible correction factor as a function of Galactic latitude that is broadly applicable to extragalactic DM observations beyond the simulations employed in this work.

\subsection{Removing Underlying Model Components}
\begin{figure}
    \centering
    \includegraphics[scale=0.32]{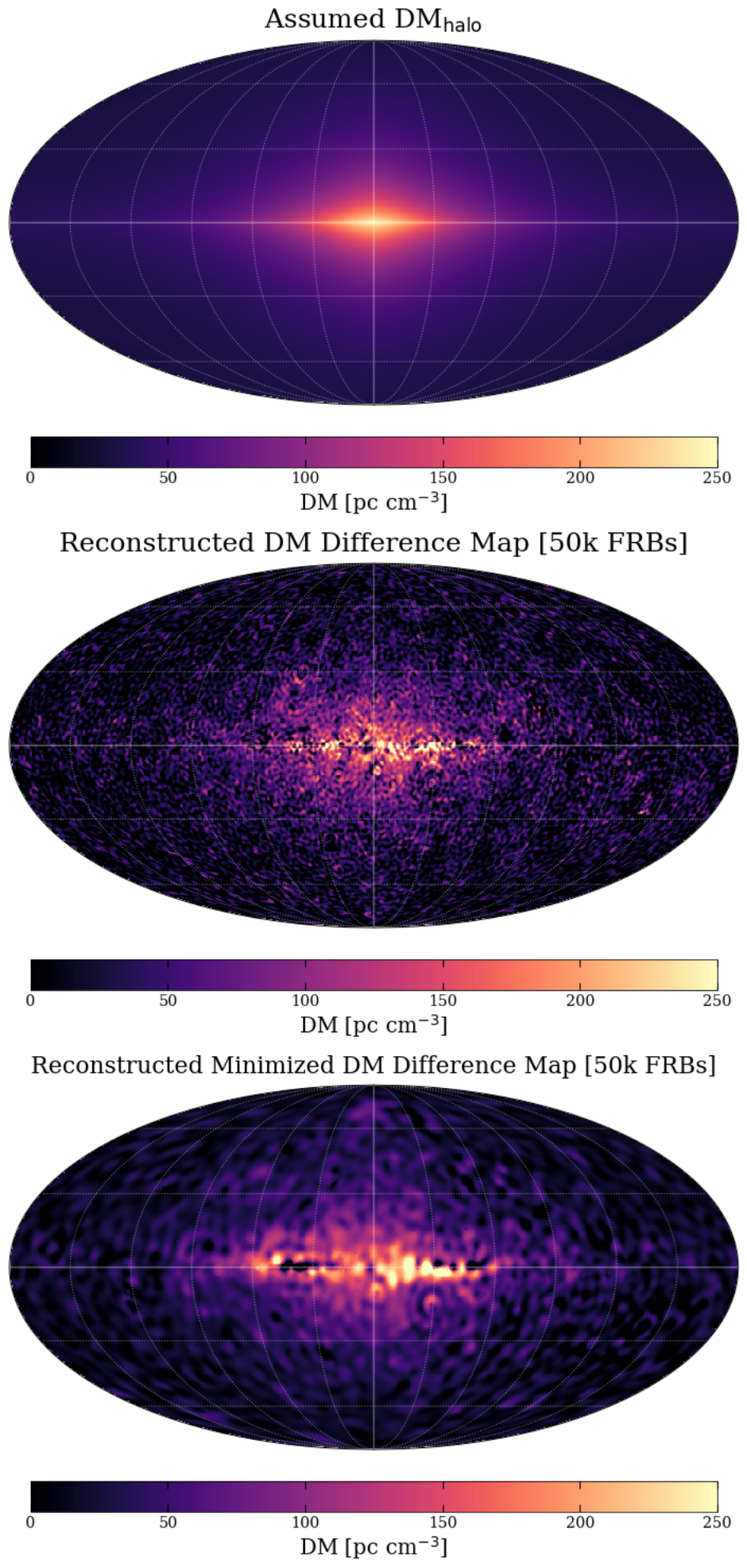}
    \caption{(Top) DM contribution from the halo component of our $n_e$ model (YT20) generated from {\tt hammurabiX}. (Middle) Difference map created by subtracting the reconstructed DM map without the YT20 halo contribution from the reconstructed result when assuming both YMW16 and YT20 models. This map uses the largest FRB population (50000 FRBs). (Bottom) Same as the middle plot but after applying the filter to minimize extragalactic DM contributions (3072 FRBs).}
    \label{fig:maps_nonehalo}
\end{figure}

To assess whether the reconstruction results are robust against changes in the underlying model, we redo the analysis in the previous section without including the YT20 halo model in our input model. The first panel of Figure \ref{fig:maps_nonehalo} depicts the expected DM contribution from this halo model alone. To generate the map in the second panel, we subtract the reconstruction results without YT20 from those that included both YT20 and YMW16 $n_e$ models. The difference in the subsequent panels of Figure \ref{fig:maps_nonehalo} clearly illustrates that the reconstruction is able to identify the change in model, particularly near the GC. Farther out in the halo, the difference map becomes dominated by random pixel-to-pixel variation in the extragalactic component between the two runs. When the extragalactic contribution is minimized in the final panel, we are able to identify the missing halo structure over a broader area of the sky. In this difference map, we encounter lower resolution structure than in the middle panel and this is due to the vastly reduced number of total data points used in the reconstruction (i.e. from 50000 FRBs in the middle panel to 3072 in the bottom panel).

\begin{figure}
    \centering
    \includegraphics[scale=0.32]{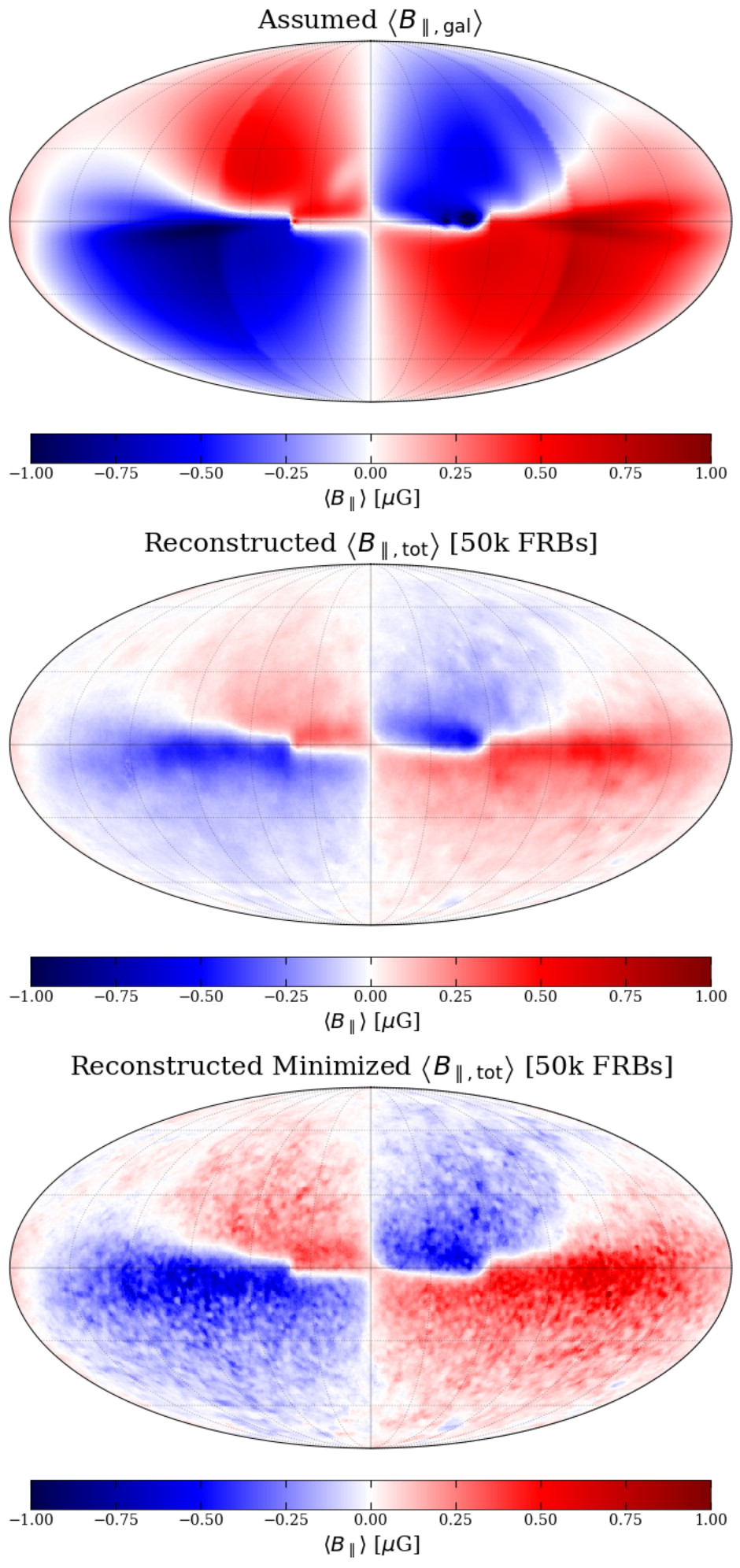}
    \caption{(Top) Expected $\left<B_{\parallel}\right>$ from {\tt hammurabiX} when excluding the poloidal X-shaped halo from the JF12 model. Reconstructed $\left<B_{\parallel}\right>$ maps with 50000 simulated FRBs (middle) and the 3072 DM-filtered data points (bottom) are also plotted.}
    \label{fig:maps_noxhalo}
\end{figure}

Likewise, we also test whether our algorithm can discern a small change in the GMF model by removing the X-shaped halo component from JF12. Figure \ref{fig:maps_noxhalo} depicts the expected $\left<B_{\parallel}\right>$ without the X-shaped halo (top), and the reconstructed $\left<B_{\parallel}\right>$ before (middle) and after (bottom) applying the DM filter. While removing this feature does not impact the magnitude of $\left<B_{\parallel}\right>$ significantly, the large-scale structure, such as the location of reversals in the $\left<B_{\parallel}\right>$ direction, is noticeably different from the first row of Figure \ref{fig:maps_gal}. Hence, we elect to directly compare to the reconstructed maps rather than analyzing difference maps. Again, we see that in both cases (full 50000 FRB sample and the DM-filtered sample), we can identify the change in model and see clear structural differences from the $\left<B_{\parallel}\right>$ reconstructions in Figure \ref{fig:maps_standard} (which includes the X-shaped halo in the GMF). As expected, the filtered data set more accurately estimates $\left<B_{\parallel}\right>$ while the full data set more strongly underestimates $\left<B_{\parallel}\right>$.

\section{Discussion} \label{discussion}

\subsection{Constraining extragalactic DM} \label{sec:dm_corr}
As we saw in Section \ref{results}, the most prominent impediment in reconstructing Galactic structure in the DM and $\left<B_{\parallel}\right>$ sky is the large extragalactic contributions to DM from FRB host galaxies and the IGM. So far, we have applied a filter to remove high DMs in equal-area portions of the sky to help mitigate this problem. Since our simulated lines of sight are roughly uniformly distributed across the sky, applying this filter allows us to effectively extract a subset of FRBs with the lowest total DMs. Physically, this is the equivalent of selecting a subset of the nearest FRBs, distributed in a sphere around the observer. The reconstruction results obtained from using this subset as an input, therefore, provide an upper limit to the total Galactic DM contribution across the sky. However, this upper limit is not ideal as we still cannot isolate the Galactic structure from the remaining extragalactic DM. To tackle this issue in greater detail, we analyze the mean $\left<B_{\parallel}\right>$, observed-to-expected $\left<B_{\parallel}\right>$ fraction and absolute difference from the corresponding {\tt hammurabiX} observable. We break the maps down into latitude strips with a width of 5 deg and sample the entire longitude range at intervals of 30 deg. A full collection of these results is presented in Appendix \ref{lon_lat_analysis} for the standard set of assumed input models (YMW16, YT20, and JF12) seen in Figure \ref{fig:maps_standard}.

\begin{figure*}
    \centering
    \includegraphics[scale=0.400]{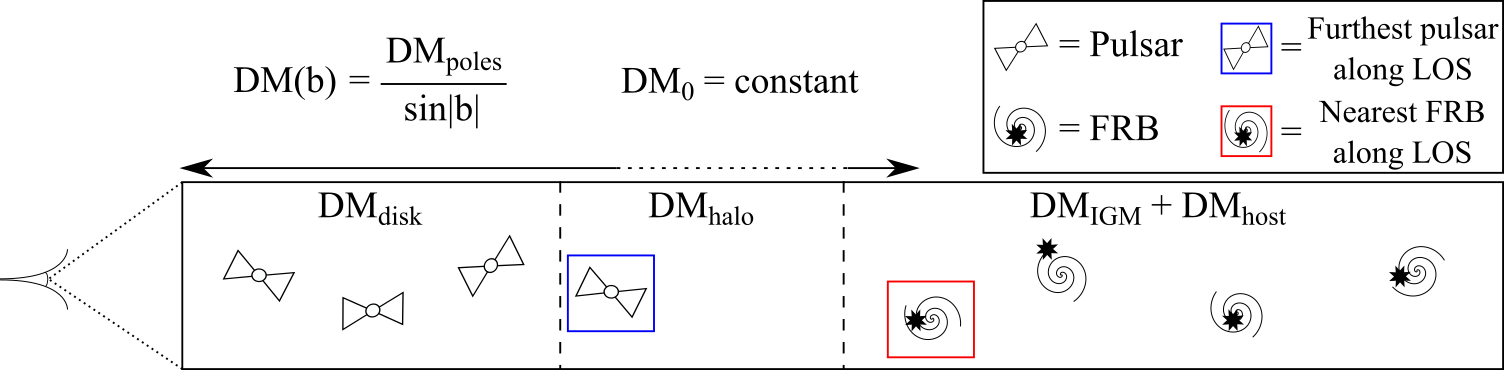}
    \caption{A simple illustration depicting the various components of DM contributions along a given LOS through the MW, as probed by pulsars and FRBs. The observed DM is comprised of contributions from the MW disk ($\mathrm{DM}_{\mathrm{disk}}$), MW halo ($\mathrm{DM}_{\mathrm{halo}}$), IGM ($\mathrm{DM}_{\mathrm{IGM}}$), and the FRB host galaxy ($\mathrm{DM}_{\mathrm{host}}$). Pulsars are depicted as circles with two symmetrical cones and FRBs and their host galaxies are represented by black bursts within a spiral; the farthest pulsar and nearest FRB are highlighted as they provide a lower and upper limit on the total Galactic DM along the LOS, respectively. Dashed lines are used to show the boundary between the MW disk, halo, and the IGM and arrows are used to indicate the physical regions for which our model (equation \ref{eq:dm_model}) is estimating DM contributions. The pulsar-based correction models everything beyond the furthest pulsar as a Galactic latitude independent extragalactic DM contribution, while the FRB-based correction includes more of the halo in the latitude dependent Galactic DM contribution. Not all of the Galactic halo is included in the latitude dependent term, even by the FRB-based correction, and the dotted section of the arrows signifies the ambiguity regarding the exact extent to which the halo is classified as Galactic DM contribution. Note that most pulsars are located within the MW disk but there are some located in the MW halo, similar to the example depicted in this diagram.}
    \label{fig:drawn_los}
\end{figure*}

Figure \ref{fig:drawn_los} shows a visualization of the different DM contributions along a given LOS through our Galaxy and the information provided by using pulsars and FRBs as probes of the Galactic DM. In this section, we aim to derive a correction factor that can be applied to the data to  distinguish between Galactic and extragalactic DM contributions. To do this, we model the total observed DM along a given LOS $\mathrm{DM}_{\mathrm{obs}}$ as a latitude dependent term $\mathrm{DM}(b)$, which encapsulates the Galactic DM contribution, plus a latitude independent term $\mathrm{DM}_0$, which represents the largely isotropic extragalactic contribution. A number of studies have previously examined the vertical structure of the ISM using DM observations from pulsars \citep[e.g.][]{2008PASA...25..184G, 2020ApJ...897..124O} and found that the Galactic DM contribution (which is dominated by the ISM) falls off with latitude as 1/$\mathrm{sin}|b|$, taking a minimum value towards the Galactic poles. Following this framework, we model $\mathrm{DM}_{\mathrm{obs}}$ as:
\begin{equation}
\mathrm{DM}_{\mathrm{obs}} = \mathrm{DM}(b) + \mathrm{DM}_0 = \frac{\mathrm{DM_{poles}}}{\mathrm{sin}|b|} + \mathrm{DM}_0~\mathrm{pc}~\mathrm{cm}^{-3}\,, \label{eq:dm_model}
\end{equation}
where $\mathrm{DM}_{\mathrm{poles}}$ is the expected DM contribution from all disk-like Galactic components along the LOS to the Galactic poles. To correct for the extragalactic DM along a given LOS ($\mathrm{DM}_i$), we define a correction factor $F = \mathrm{DM}(b) / \mathrm{DM}_{\mathrm{obs}}$ such that:
\begin{equation}
F = \frac{\mathrm{DM}(b)}{\mathrm{DM}(b) + \mathrm{DM}_0}\,;
\label{eq:corr_fac}
\end{equation}
\begin{equation}
\mathrm{DM}_{i,\mathrm{corr}} = F \mathrm{DM}_i\,, \label{eq:dm_corr}
\end{equation}
where $\mathrm{DM}_{i,\mathrm{corr}}$ is the corrected DM along the LOS.

One method of applying this correction factor is to use pulsar data in estimating the value of $\mathrm{DM}_{\mathrm{poles}}$ and then fitting DM data to equation \ref{eq:dm_model} to determine the average value of $\mathrm{DM}_0$ over the sky. Hereafter, this approach will be referred to as the pulsar-based correction. Note that the pulsar-based correction is limited by the most distant pulsar in a data set and cannot probe any additional Galactic DM contributions from the halo beyond that point. When applying this method, we follow the procedure detailed by \cite{2020ApJ...897..124O}, using the same set of pulsar data (see their Table 1), except with DMs computed according to our assumed Galactic $n_e$ models (YMW16 and YT20). The resulting least squares best fit estimates for equation \ref{eq:dm_model} are $\mathrm{DM}_{\mathrm{poles}} = 33.6 \pm 4.2~\mathrm{pc}~\mathrm{cm}^{-3}$ and $\mathrm{DM}_0 = 150.7 \pm 6.3~\mathrm{pc}~\mathrm{cm}^{-3}$. Applying this correction factor to the subset of data that has been filtered to minimize extragalactic DMs then provides a lower limit to the total Galactic DM contribution. Namely, it provides an estimate of the DM contribution from the ISM. With regards to the GMF, applying this correction factor will give us an upper limit for the total Galactic $\left|\left<B_{\parallel}\right>\right|$.

Alternatively, we can also use the FRB data itself when computing a correction factor by fitting equation \ref{eq:dm_model} directly to our simulated data that have been filtered to minimize extragalactic DMs, leaving both $\mathrm{DM}_{\mathrm{poles}}$ and $\mathrm{DM}_0$ as free variables. Hereafter, this approach will be referred to as the FRB-based correction. Effectively, the FRB-based correction is directly separating out the DM data into Galactic latitude dependent and independent terms to distinguish Galactic and extragalactic DM contributions. The least squares best fit estimates for the FRB-based correction are $\mathrm{DM}_{\mathrm{poles}} = 41.7 \pm 5.0~\mathrm{pc}~\mathrm{cm}^{-3}$ and $\mathrm{DM}_0 = 135.6 \pm 11.1~\mathrm{pc}~\mathrm{cm}^{-3}$. Since FRBs are extragalactic sources and probe the entirety of the MW, the FRB-based correction is able to better estimate the Galactic contributions from the MW halo than the pulsar-based correction. However, we expect the furthest reaches of the MW halo to be an extended spherical $n_e$ halo \citep{2020ApJ...888..105Y}, which will add a roughly constant and isotropic DM contribution across the sky, and be indistinguishable from the extragalactic contributions. Therefore, while the FRB-based correction provides a better estimate of the total Galactic DM contribution than the pulsar-based estimate, it is still does not exactly reconstruct the input Galactic DM and $\left|\left<B_{\parallel}\right>\right|$.

\begin{figure}
    \centering
    \includegraphics[scale=0.36]{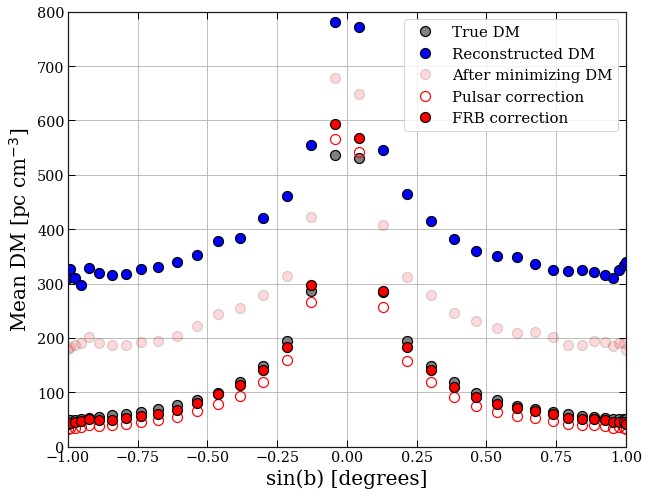}
    \caption{Mean DM over 0 deg < $\ell$ < 360 deg plotted as a function of the sine of the Galactic latitude $b$. The input Galactic DM contributions from {\tt hammurabiX} are plotted in grey. Blue points represent the full set of 50000 simulated FRB observations and evidently contain significant extragalactic contributions. Filtering the data to remove high DM lines of sight, as mentioned in Section \ref{results}, reduces the extragalactic contribution and provides an upper limit to the Galactic DM; these points are plotted as light red. Further, applying the pulsar-based DM correction estimates the $\mathrm{DM}_{\mathrm{ISM}}$ and places a lower limit on the total Galactic DM$-$as depicted by the hollow red points. Instead, applying the FRB-based correction by allowing both $\mathrm{DM}_{\mathrm{poles}}$ and $\mathrm{DM}_{\mathrm{xgal,mean}}$ to be free parameters, produces a much closer fit to the input Galactic DM and is plotted in dark red. Note that the corrections applied to obtain the hollow red and dark red points are only valid for $|b|$ > 10~deg (i.e. $|\rm{sin}(b)| \gtrsim 0.17$).}
    \label{fig:corrfac}
\end{figure}

\begin{figure}
    \centering
    \includegraphics[scale=0.195]{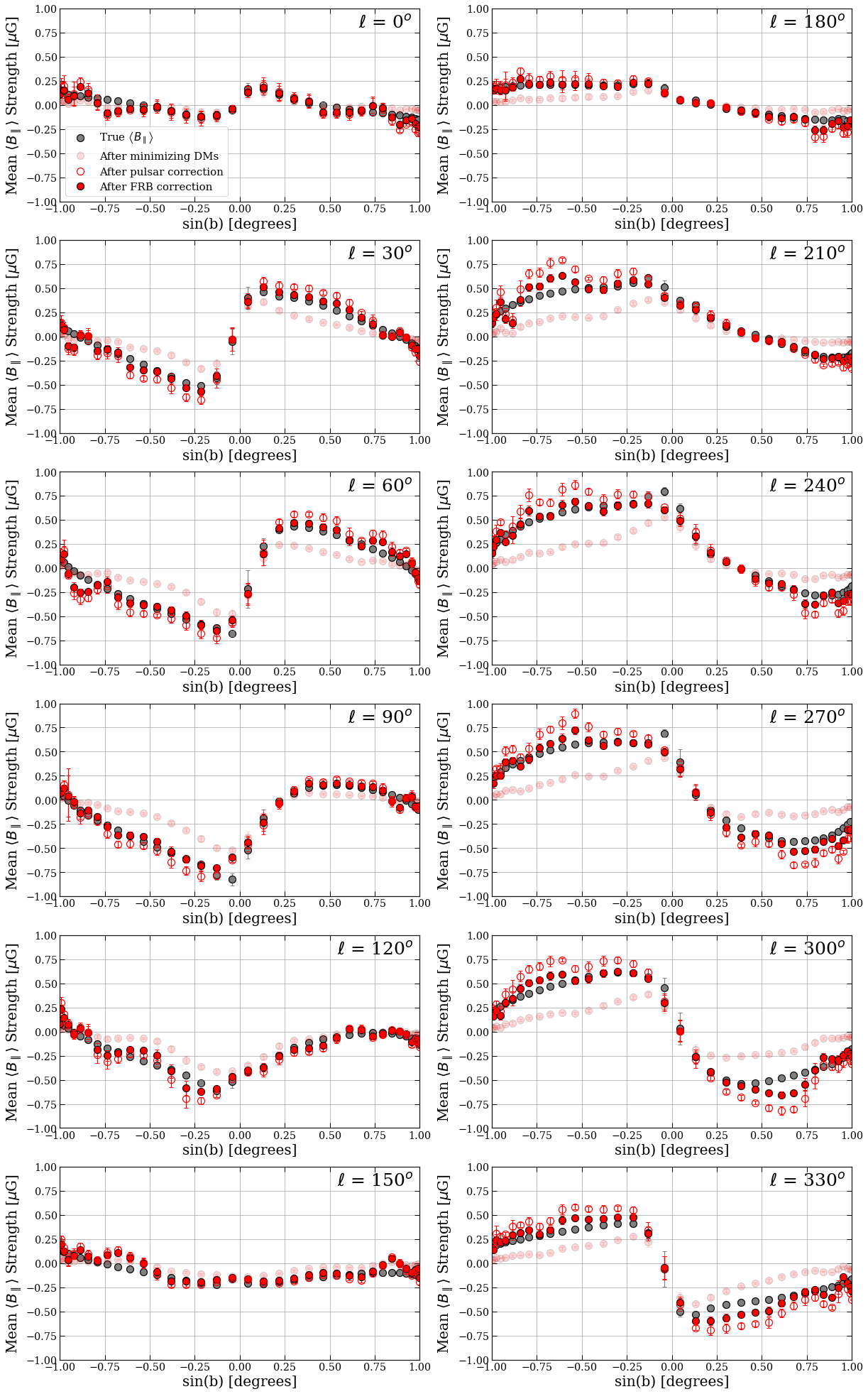}
    \caption{A comprehensive look at the effects of applying the pulsar-based and FRB-based DM corrections, shown in Figure \ref{fig:corrfac} to $\left<B_{\parallel}\right>$ at various longitudes cut. Each sub-plot is offset by 30 deg in Galactic longitude $\ell$ and marker selection is identical to Figure \ref{fig:corrfac}$-$with light red (hollow red) points now representing a lower limit (upper limit) in $\left<B_{\parallel}\right>$.}
    \label{fig:corrfac_b}
\end{figure}

In both the pulsar-based and FRB-based corrections, we note that the 1/$\mathrm{sin}|b|$ assumption from equation \ref{eq:dm_model} breaks down as $b \rightarrow$ 0. As a result, our correction factor becomes unreliable at very low Galactic latitudes and, subsequently, we only apply corrections for $|b| > 10~\mathrm{deg}$. Figure \ref{fig:corrfac} summarizes the various steps taken to disentangle the Galactic and extragalactic DM contributions and shows that the FRB-based correction provides the better estimate of the input Galactic DM at $|b| > 10~\mathrm{deg}$. Specifically, the combination of applying a filter to remove high DMs in equal-area portions of the sky and then applying the FRB-based correction (via equations \ref{eq:dm_model}, \ref{eq:corr_fac}, and \ref{eq:dm_corr}) reduces the reconstructed DM by an average of $272.9 \pm 8.4~\mathrm{pc}~\mathrm{cm}^{-3}$ for $|b|$ > 10~deg and the absolute difference between this corrected DM and the input model DM is on average $\lesssim 6.1 \pm 2.4~\mathrm{pc}~\mathrm{cm}^{-3}$ for $|b|$ > 10~deg. Applying the reciprocal of equation \ref{eq:corr_fac} to $\left<B_{\parallel}\right>$ improves the accuracy of our magnetic field reconstruction as well. Figure \ref{fig:corrfac_b} depicts the results of applying our pulsar-based and FRB-based corrections to the observed $\left<B_{\parallel}\right>$ at slices taken at intervals of 30~deg in Galactic longitude. For each of the panels in Figure \ref{fig:corrfac_b}, the absolute difference between the FRB-based correction of $\left<B_{\parallel}\right>$ and input model $\left<B_{\parallel}\right>$ is on average $\lesssim 0.06 \pm 0.04~\mu$G for $|b|$ > 10~deg.

To further establish that the correction factor can be applied beyond the standard assumed model examined above, we repeat the same process after removing the X-shaped halo from the GMF model. Applying the same corrections, we are again able to recover DM and $\left<B_{\parallel}\right>$ structure that better resembles the expected Galactic observables. In this case, following the pulsar-based correction gives us a best fit of $\mathrm{DM}_{\mathrm{poles}} = 33.6 \pm 4.2~\mathrm{pc}~\mathrm{cm}^{-3}$ and $\mathrm{DM}_0 = 152.0 \pm 6.3~\mathrm{pc}~\mathrm{cm}^{-3}$, and following the FRB-based correction gives $\mathrm{DM}_{\mathrm{poles}} = 42.3 \pm 4.9~\mathrm{pc}~\mathrm{cm}^{-3}$ and $\mathrm{DM}_0 = 135.8 \pm 10.9~\mathrm{pc}~\mathrm{cm}^{-3}$. Taking the absolute difference between our best correction (FRB-based) and the input DM ($\left<B_{\parallel}\right>$), we find that on average they differ by $\lesssim 4.8 \pm 1.8~\mathrm{pc}~\mathrm{cm}^{-3}$ ($\lesssim 0.07 \pm 0.06~\mu$G) for $|b|$ > 10~deg. See Appendix \ref{lon_lat_analysis} for the full set of corresponding plots.

\subsection{Application to Observed Data \& Caveats}
Throughout this paper, we have presented results based on a simplified model. Here, we discuss application this work has to current and future FRB observations. Firstly, these simulated observations are derived from models of various components of the MW. While these models are useful for describing the Galactic structure on large scales, we expect there to be a great deal more turbulence and small scale structure in observations (e.g. compare the modelled RM$_{\rm{gal}}$ from Figure \ref{fig:maps_gal} to the results presented by \citealp{2012A&A...542A..93O} and \citealp{2020A&A...633A.150H}, which are derived from observations). This has implications on the expected power spectra for both the RM and DM skies. In this work, our inference algorithm has expected rather steep power spectra, corresponding to sky maps dominated by large structures such as the Galactic disc. For a real data application, this expectation (i.e. the prior on the power spectrum slope) would have to be adapted. 

The simulated FRB populations used in this work are also largely simplified versions of observed FRBs. For instance, in the first CHIME/FRB catalogue \citep{2021ApJS..257...59A}, the observed sky distribution of sources only extends down to DEC $\sim$ $-$11~deg and is not uniformly distributed. This is due to the transiting observing mode used for CHIME; the distribution of FRBs detected is most dense around the celestial north pole and decreases away from it. This is a major point to note because, in our simulated observations, we had a fairly evenly distributed FRB population over the whole sky which allowed for the joint inference to parameterize structure even with only 1000 FRBs. In reality, we may require many more observations to achieve a sufficient minimum source density over the entire observable sky area. However, it should be noted that DM observations from several FRB surveys can be combined into a single catalogue and input into the joint inference algorithm (similar to RM catalogue utilized in this work).

We note that while the extragalactic DM contribution in our model is isotropic (which we expect observationally on large scales), in reality it may be structured over small scales. This would be due to contributions from intervening extragalactic sources along the LOS (e.g. M31, the Magellanic Clouds, and LGM) \citep[e.g.][]{2019MNRAS.485..648P}, which is not entirely deleterious as it allows for the study of DM structure inherent to these objects \citep[e.g.][]{2021arXiv210713692C}. However, for the purposes of studying the MW, these contributions will require a more careful handling of the data or the inference model to constrain.

Another consideration is that the current CHIME/FRB catalogue has relatively low positional accuracy, with positional uncertainties up to $\sim 0.5~\mathrm{deg}$. These positional uncertainties give us a lower limit on the size of Galactic structure that can be resolved using our joint-inference reconstruction method. However, the mean spacing of our FRBs is larger than the typical FRB positional uncertainties after applying the DM filtering criterion, and thus the former dominates the scale of resolvable Galactic structures. Nonetheless, to ensure that these uncertainties do not significantly affect the joint-inference reconstruction, we repeated our analysis (Section \ref{results}) after shifting each of the simulated FRBs in our sample by a randomly drawn RA and DEC uncertainty from the CHIME/FRB catalogue. The typical on-sky offset of each simulated FRB from its original position was $\sim 0.26~\mathrm{deg}$. After applying the DM filtering criterion and generating the reconstructed map using the shifted data set, the average percent difference in DM for a given pixel was $\sim 4-5\%$ compared to the original reconstruction. The maximum percent difference in DM for a given pixel was $\sim 40\%$. Further, the estimates for $\mathrm{DM}_{\mathrm{poles}}$ and $\mathrm{DM}_0$ from the reconstruction using shifted FRBs only varied $\sim 2-3\%$ from the original values ($\mathrm{DM}_{\mathrm{poles}} = 41.7 \pm 5.0~\mathrm{pc}~\mathrm{cm}^{-3}$ and $\mathrm{DM}_0 = 135.6 \pm 11.1~\mathrm{pc}~\mathrm{cm}^{-3}$) and were in agreement with the quoted uncertainties. Therefore, we find that artificially including FRB positional uncertainties typical of the \cite{2021ApJS..257...59A} catalogue does not significantly change the reconstruction results and major conclusions of our analysis.
We also note that we will have better FRB localizations with future FRB surveys, which are described in further detail below, and therefore we expect the typical positional accuracy of FRBs to improve in the near future.

As the existing catalog of FRBs grows, the aspects of future FRB observations that will be particularly valuable to developing this reconstruction method include: (i) a large number of nearby FRBs (since low DM FRBs are more informative to this method than distant, high DM FRBs), (ii) all-sky coverage, and (iii) associated redshift information (allowing us to estimate extragalactic DM contributions). The premier survey to obtain such a data set currently, and in the near future, is CHIME/FRB with its outrigger facilities which will supplement the high detection rate and daily all-hemisphere coverage of CHIME/FRB with better on-sky localization, thus leading to more associated redshift estimates \citep[for details regarding the CHIME outrigger systems and localization of FRBs through Very Long Baseline Interferometry, see][]{2022AJ....163...48M, 2022AJ....163...65C}. On longer time scales, we expect the Hydrogen Intensity and Real-time Analysis Experiment \citep[HIRAX,][]{HIRAX} and the Canadian Hydrogen Observatory and Radio-transient Detector \citep[CHORD,][]{CHORD} to come online and have similar, or higher, FRB rates than CHIME/FRB. However, HIRAX and CHORD will have a much smaller daily sky coverage than CHIME/FRB and require manual repointing, so the timescale or capacity for $\gtrsim \pi\mathrm{sr}~\mathrm{deg}$ sky coverage by either survey alone is uncertain. The Deep Synoptic Array-2000 (DSA-2000) is expected to see first light in 2026, covering $\sim 3\pi\mathrm{sr}~\mathrm{deg}$ every few months and detecting/localizing FRBs at a rate of $\sim 10^4~\mathrm{year}^{-1}$ \citep{DSA2000}. In addition, the new coherent search system on the Commensal Real-Time Australia Square Kilometer Array Pathfinder Fast-Transient survey \citep[CRAFT,][]{CRAFT} is expected to provide $\sim 1.5$ localized FRBs per day. CRAFT is a commensal search with the Australian Square Kilometre Array Pathfinder \citep[ASKAP,][]{ASKAP} survey science projects \citep{ASKAPsci} and will cover the entire southern sky in approximately 5 years. With the reconstruction technique described in Section \ref{inference}, we can input a combined FRB catalogue comprising data from multiple surveys, allowing us to exploit the complementary sky coverage of different surveys (e.g. CHIME/FRB, CHORD, and the DSA-2000 in the northern hemisphere, and HIRAX and CRAFT in the southern hemisphere).

While we expect the catalog of FRBs to increase greatly in the near future, the number of radio galaxy RMs will proliferate as we get data releases from large sky radio surveys such as VLASS and POSSUM. In the next 5 years, the radio galaxy RM catalog will increase from $\sim 50000$ to $\sim 10^6$ sources \citep{2020Galax...8...53H}. Therefore, as is currently the case (see Section \ref{intro}), the number of radio galaxy RMs will still far exceed the number of FRB RMs in the near future.

\subsection{Interpreting GMF Results \& Exceptions}
Throughout this study we have been focused on $\left<B_{\parallel}\right>$; it is important to contextualize what exactly $\left<B_{\parallel}\right>$ describes relative to $B_{\parallel}$ or the full GMF and to remember the assumptions that went into deriving equation \ref{eq:B} and the resulting reconstructions. $\left<B_{\parallel}\right>$ provides the electron density weighted average magnetic field strength and net direction$-$this quantity is not necessarily equivalent to $B_{\parallel}$ and only provides the averaged $B_{\parallel}$ strength and net orientation along the entire distance between the observer and the emitting object. Hence, $\left<B_{\parallel}\right>$ measurements do not provide us with knowledge of $B_{\parallel}$ at any specific distance along the LOS. A reconstruction of the $\left<B_{\parallel}\right>$ sky also does not provide us with full information about the GMF. To get a more complete picture of the GMF, $\left<B_{\parallel}\right>$ must be analyzed alongside the plane-of-sky magnetic field $B_{\perp}$.

Recall that, for equation \ref{eq:B}, we made the implicit assumption that the electron density $n_e$ and $B_{\parallel}$ are uncorrelated$-$however this may not always be a good assumption. Previous studies \citep[e.g.][]{2003A&A...411...99B, 2021MNRAS.502.2220S} have shown that measurements of $\left<B_{\parallel}\right>$ using equation \ref{eq:B} may be underestimated or overestimated in regions where $n_e$ and $B_{\parallel}$ are negatively or positively correlated, respectively. Specifically, \cite{2021MNRAS.502.2220S} showed that $n_e$ and $B_{\parallel}$ are largely uncorrelated over kpc scales but may break down at sub-kpc scales. Since this work is primarily focused on reconstructing $\left<B_{\parallel}\right>$ over large Galactic scales, equation \ref{eq:B} provides a valid estimate of the true underlying $B_{\parallel}$.

This study focuses on the halo rather than the disk. The halo structure is likely much smoother than the disk since in the disk we expect to find more dense patches of electrons, e.g., due to star formation regions. We expect that equation \ref{eq:B} is more likely to be true in a region such as the halo, where the electron density is smoothly varying.

\section{Conclusions} \label{conclusion}
We present a new method of reconstructing all-sky $\left<B_{\parallel}\right>$ information for the Galactic magnetic field (GMF) using a rotation measure$-$dispersion measure (RM$-$DM) joint inference algorithm based on information field theory \citep{2009PhRvD..80j5005E, 2019AnP...53100127E} and demonstrate its effectiveness on simulated line of sight (LOS) observations to radio galaxies and fast radio burst (FRB) populations. The joint inference is able to constrain DM, RM, and $\left<B_{\parallel}\right>$ along a given LOS by exploiting correlations of nearby data points and each sky map is parameterized as a non-linear combination of Gaussian sky maps, assuming $\mathrm{RM}_{\mathrm{gal}}$ and $\mathrm{DM}_{\mathrm{gal}}$ are modelled by equations \ref{eq:rm_inference_model} and \ref{eq:dm_inference_model}, respectively.

Given only Galactic inputs, which serve as a baseline test but are otherwise unphysical, both the large and small scale Galactic structure of the reconstruction result is qualitatively well-matched with the input model (Figure \ref{fig:maps_gal}). While the reconstruction is hindered with the presence of extragalactic DM contributions, we detail a method of reducing and characterizing this component of DM observations (Figure \ref{fig:corrfac}) by: (i) systematically filtering out high DM observations in equal-area subsections of the sky, which provides an upper limit on Galactic DM and (ii) applying either a pulsar-based or FRB-based DM correction factor for $|b|$ > 10~deg (equations \ref{eq:dm_model}, \ref{eq:corr_fac}, and \ref{eq:dm_corr}) for an estimate of the total Galactic DM contribution.

In general, the reconstruction performs noticeably better on a larger sample of simulated FRBs (50000 FRBs) but is still able to replicate some large-scale structures, such as the high DM and $\left<B_{\parallel}\right>$ magnitude in the Galactic plane, even with only 1000 FRBs. The reconstruction is also able to identify deviations in the underlying models for both the $n_e$ distribution (removal of the \citealp{2020ApJ...888..105Y} halo model, see Figure \ref{fig:maps_nonehalo}) and GMF (removal of the \citealp{2012ApJ...757...14J} X-shaped halo, see Figure \ref{fig:maps_noxhalo}).

The results are supplemented with a discussion of the differences between our simulations and FRB observations which informs the feasibility of using current and future FRB data as a probe for large scale Galactic structure. Chief among the challenges of applying this method to FRB observations is the low number of data points that are currently publicly available and that they are not uniformly distributed over the sky. However, the rapid rate at which FRBs are being observed, especially by CHIME/FRB, is a promising step towards overcoming this hurdle. Overall, this work provides a powerful tool for studying DM, RM, and $\left<B_{\parallel}\right>$ across the entire sky with the rapidly growing FRB catalogue. In addition, there are still various avenues for further study based on this work, including:

\begin{enumerate}
    \item Incorporating extragalactic contributions, particularly for DM, directly into the joint inference algorithm as effective noise \citep[similar to the approach presented by][]{2015A&A...575A.118O, 2020A&A...633A.150H};
    \item Incorporating Galactic pulsar DM and RM measurements or Emission measures;
    \item Incorporating FRB positional uncertainties directly into the reconstruction by resampling positions in the Numerical Information Field Theory Python package (NIFTy);
    \item Designing a simulated data set that more closely resembles the sky-coverage and sensitivity of CHIME/FRB to estimate the number of FRB observations required to perform an accurate reconstruction;
    \item Testing the joint inference algorithm and correction factor (equation \ref{eq:corr_fac}) assuming other GMF or $n_e$ models. This may also include the injection of a random component to these models;
    \item Applying this method directly to FRB observations to study all-sky $\left<B_{\parallel}\right>$ and constrain the GMF structure on an unprecedented scale. This could further incorporate $B_{\perp}$ information derived from dust polarization \citep[e.g.][]{2016A&A...596A.103P} to analyze the 3D GMF.
\end{enumerate}

\section{Acknowledgements} \label{ack}
We thank Tess Jaffe and Jiaxin Wang for sharing information regarding {\tt hammurabiX} and Ziggy Pleunis, Samantha Berek, and Steffani Grondin for useful comments on the manuscript. We also thank the anonymous reviewer for careful reading of the manuscript and for providing constructive feedback. The Dunlap Institute is funded through an endowment established by the David Dunlap family and the University of Toronto. SH acknowledges funding from the European Research Council (ERC) under the European Union's Horizon 2020 research and innovation programme (grant agreement No. 772663). JLW and BMG acknowledge the support of the Natural Sciences and Engineering Research Council of Canada (NSERC) through grant RGPIN-2015-05948, and of the Canada Research Chairs program.

\section{Data Availability} \label{data_avail}
No new data were generated or analysed in support of this research. Detailed steps on how to recreate the simulated observations are laid out in section \ref{simobs}.

\bibliographystyle{mnras}
\bibliography{ref}

\appendix

\section{Summary of Reconstruction Results} \label{recon_results}
\begin{figure*}
    \centering
    \includegraphics[scale=0.315]{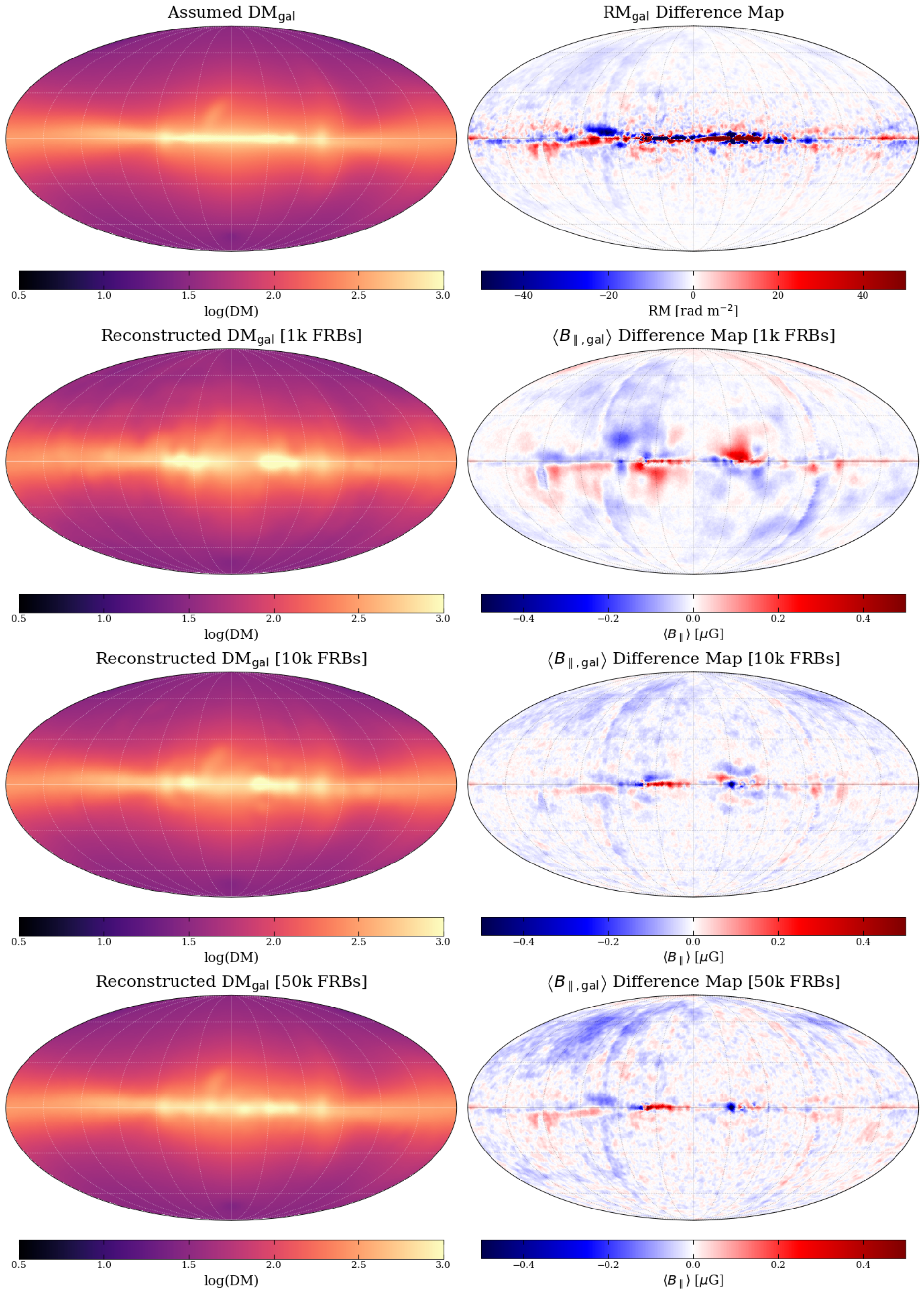}
    \caption{Supplementary Figure to Figure \ref{fig:maps_gal} presenting DM sky maps in log-space to better visualize low-DM regions away from the Galactic plane. Difference maps comparing RM and $\left<B_{\parallel}\right>$ reconstruction results to the corresponding {\tt hammurabiX} maps, seen in Figure \ref{fig:maps_gal}, are also presented in the right hand column to better highlight the small differences between the reconstruction and underlying fields.}
    \label{fig:fig4_supp}
\end{figure*}

\begin{figure*}
    \centering
    \includegraphics[scale=0.18]{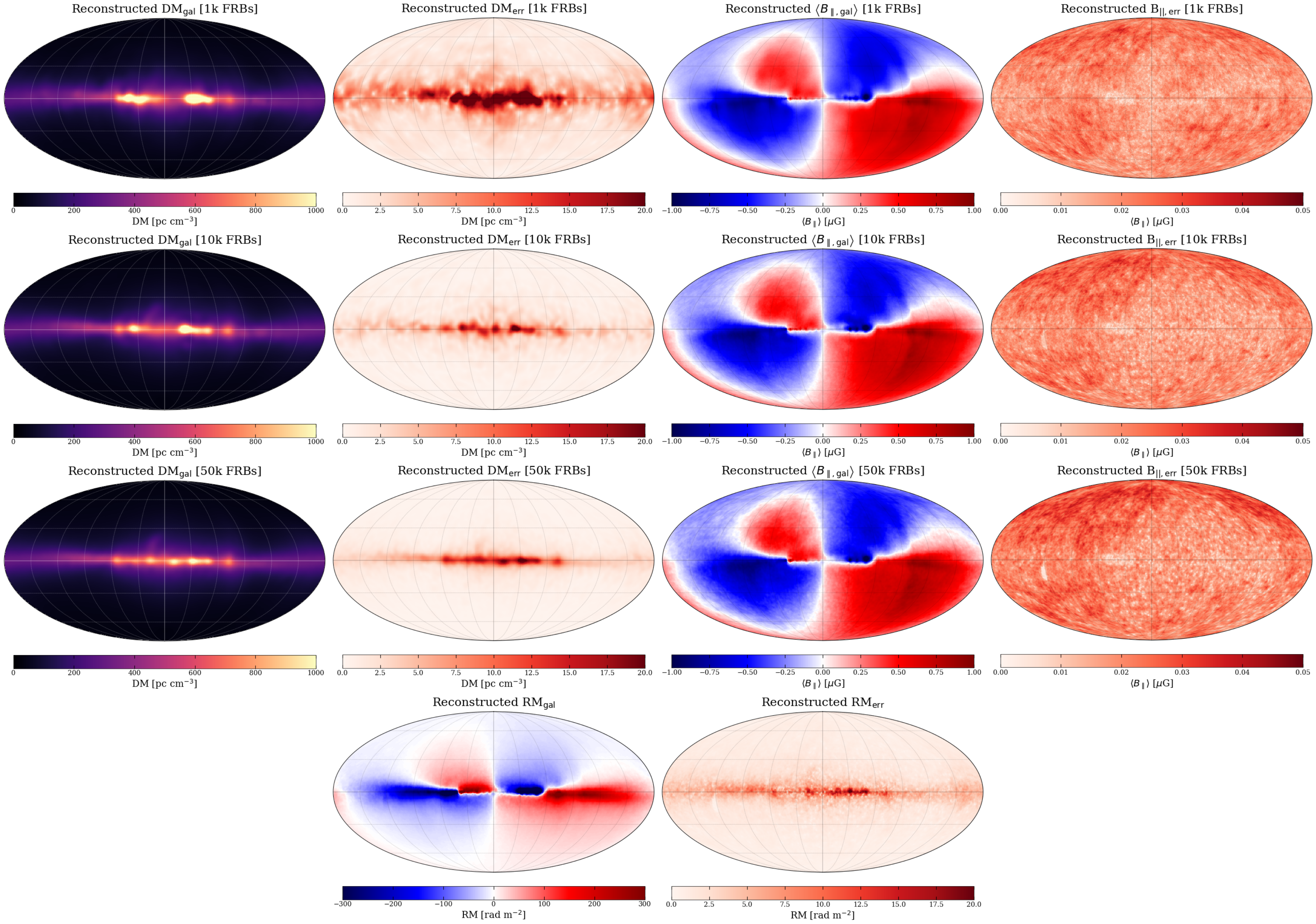}
    \caption{Full reconstruction outputs with errors corresponding to the result shown in Figure \ref{fig:maps_gal}.}
    \label{fig:all_maps_gal}
\end{figure*}

\begin{figure*}
    \centering
    \includegraphics[scale=0.18]{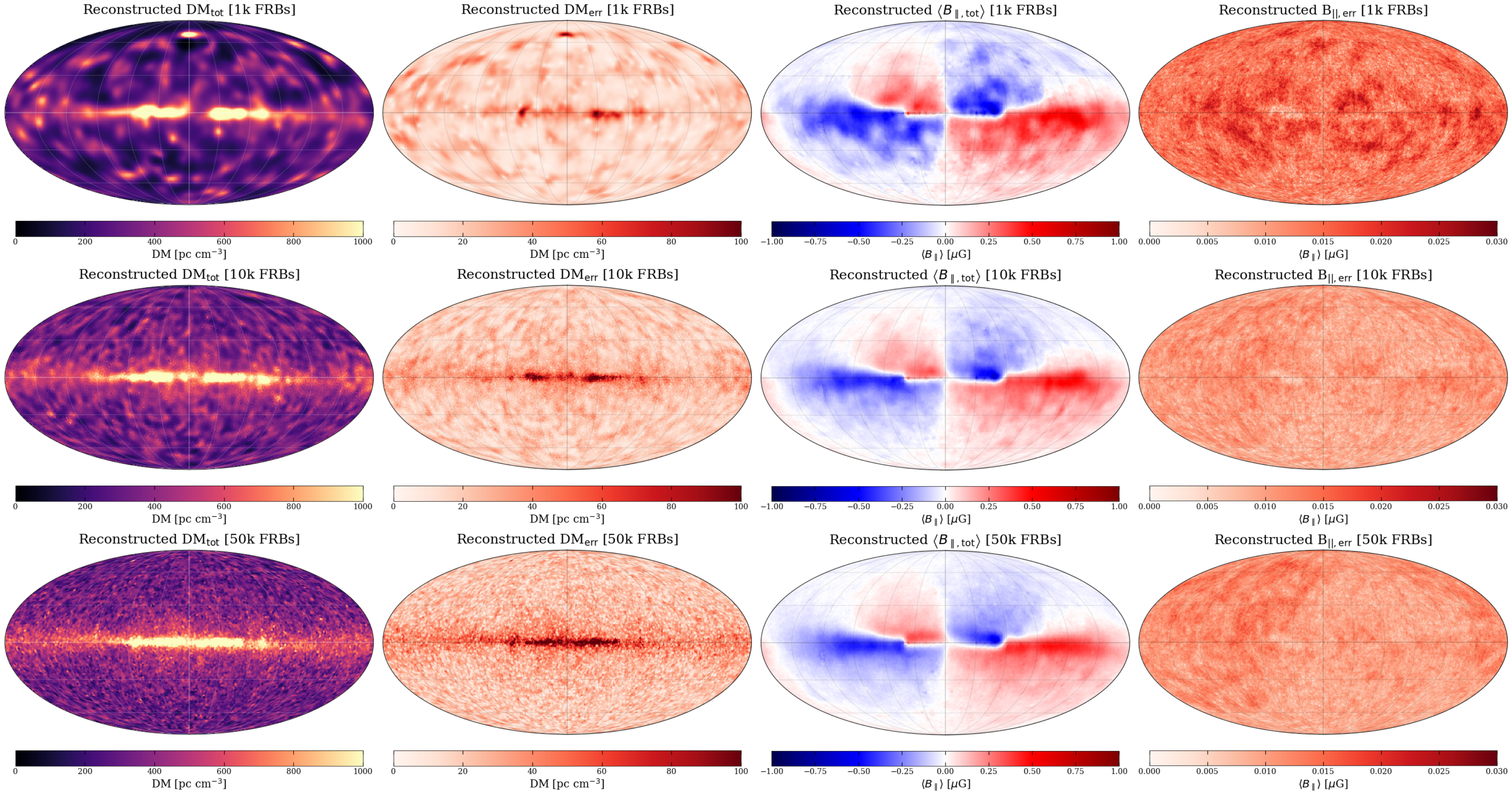}
    \caption{Identical setup to Figure \ref{fig:all_maps_gal} but corresponding to the results shown in the first two columns of Figure \ref{fig:maps_standard}.}
    \label{fig:all_maps_standard}
\end{figure*}

\begin{figure*}
    \centering
    \includegraphics[scale=0.18]{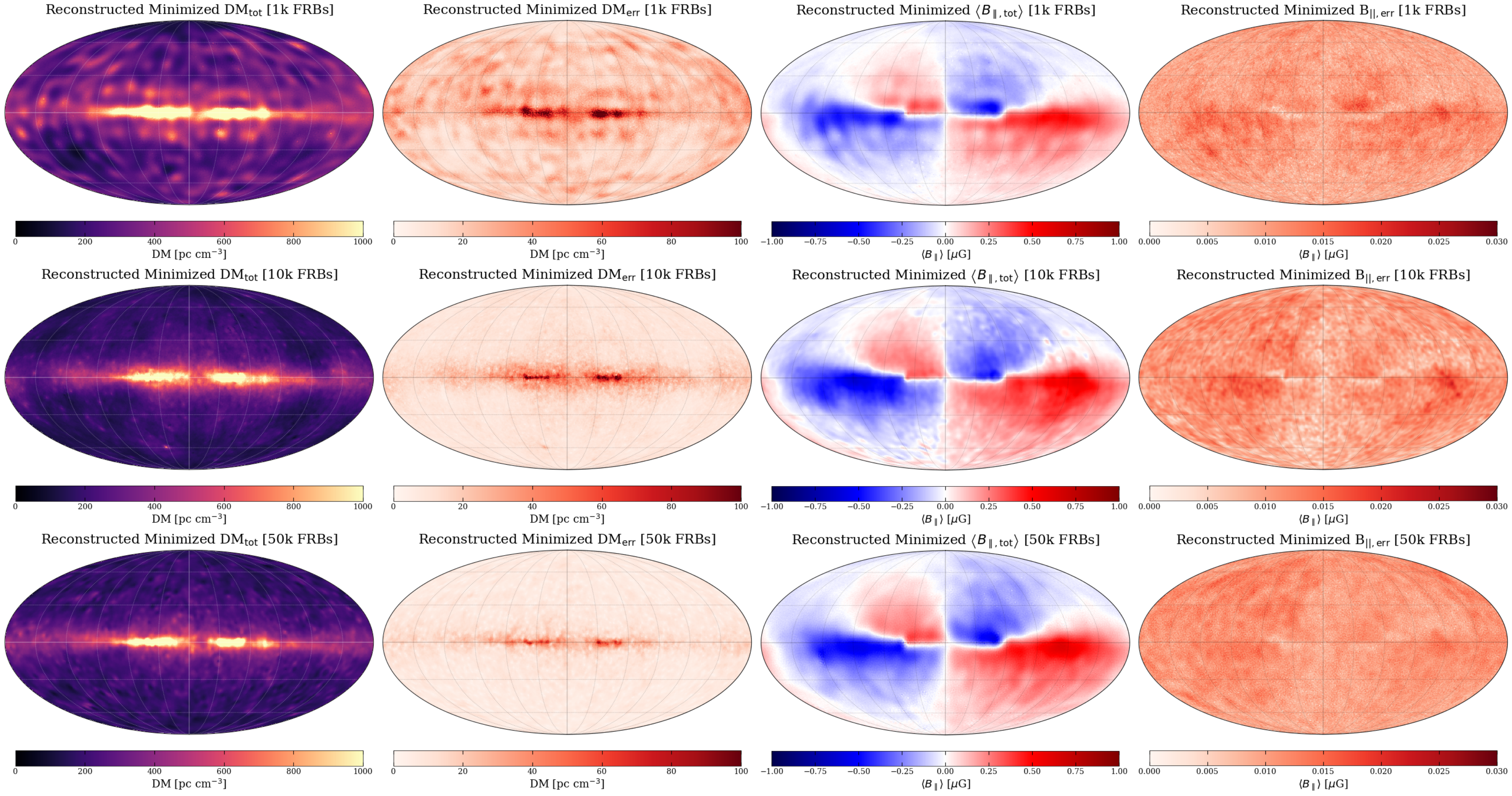}
    \caption{Identical setup to the previous two figures but corresponding to the results shown in the last two columns of Figure \ref{fig:maps_standard}.}
    \label{fig:all_maps_min}
\end{figure*}

Figure \ref{fig:fig4_supp} presents the results from Figure \ref{fig:maps_gal} in alternative ways to better visualize the differences between the reconstruction results and underlying fields. The left hand column shows the sames DM maps from Figure \ref{fig:maps_gal} but in a log base 10 scale to better see low-DM regions away from the Galactic plane. The right hand column shows difference maps between the reconstructed RM and $\left<B_{\parallel}\right>$ in Figure \ref{fig:maps_gal} and the corresponding {\tt hammurabiX} outputs. Summarized in Figures \ref{fig:all_maps_gal}, \ref{fig:all_maps_standard}, and \ref{fig:all_maps_min} are the full set of reconstructed maps with corresponding error maps. The setup of the results is identical in all three plots, starting with the largest data set of 50000 FRBs in the first row and showing the results from smaller data sets in consecutive rows. Figure \ref{fig:all_maps_gal} also has an additional row for the reconstructed RM results, which is not present in the other plots. In the case of Figure \ref{fig:all_maps_min}, recall that the data sets are much smaller after being filtered in DM space (3072, 768, and 563 FRBs respectively) but are labelled with regard to the initial size of the data set prior to the data reduction step.

\section{Correction Factor Analysis} \label{lon_lat_analysis}
\begin{figure*}
    \centering
    \includegraphics[scale=0.138]{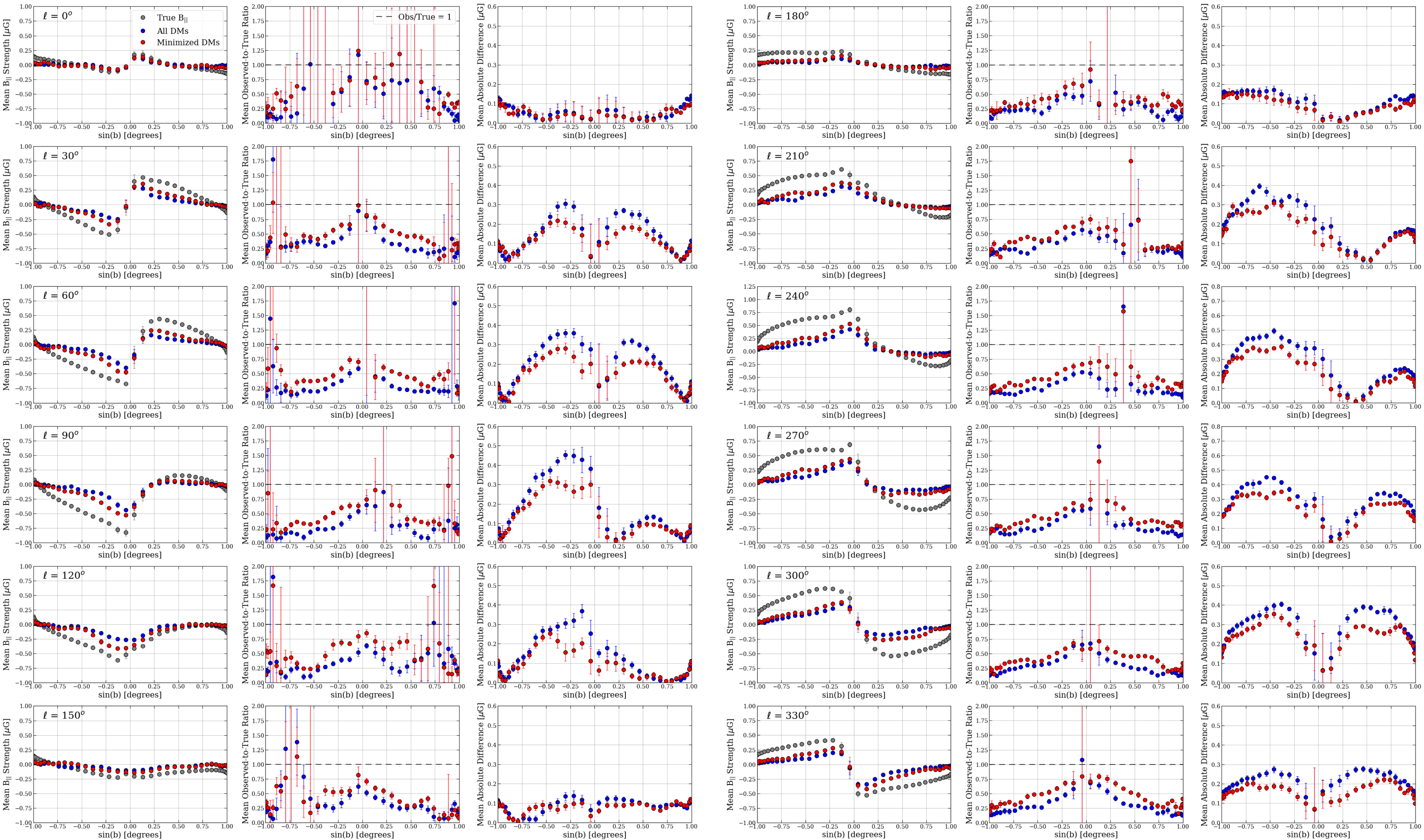}
    \caption{Analysis of the mean $\left<B_{\parallel}\right>$ as a function of Galactic latitude at 30 deg intervals in the Galactic longitude. Expected results are plotted as grey circles, and reconstruction results are in blue (full data) and red (DM filtered data). For each slice in Galactic longitude, we show the mean $\left<B_{\parallel}\right>$, observed-to-expected ratio of $\left<B_{\parallel}\right>$, and absolute difference between expected and observed results, respectively.}
    \label{fig:bpar_analysis}
\end{figure*}

\begin{figure*}
    \centering
    \includegraphics[scale=0.138]{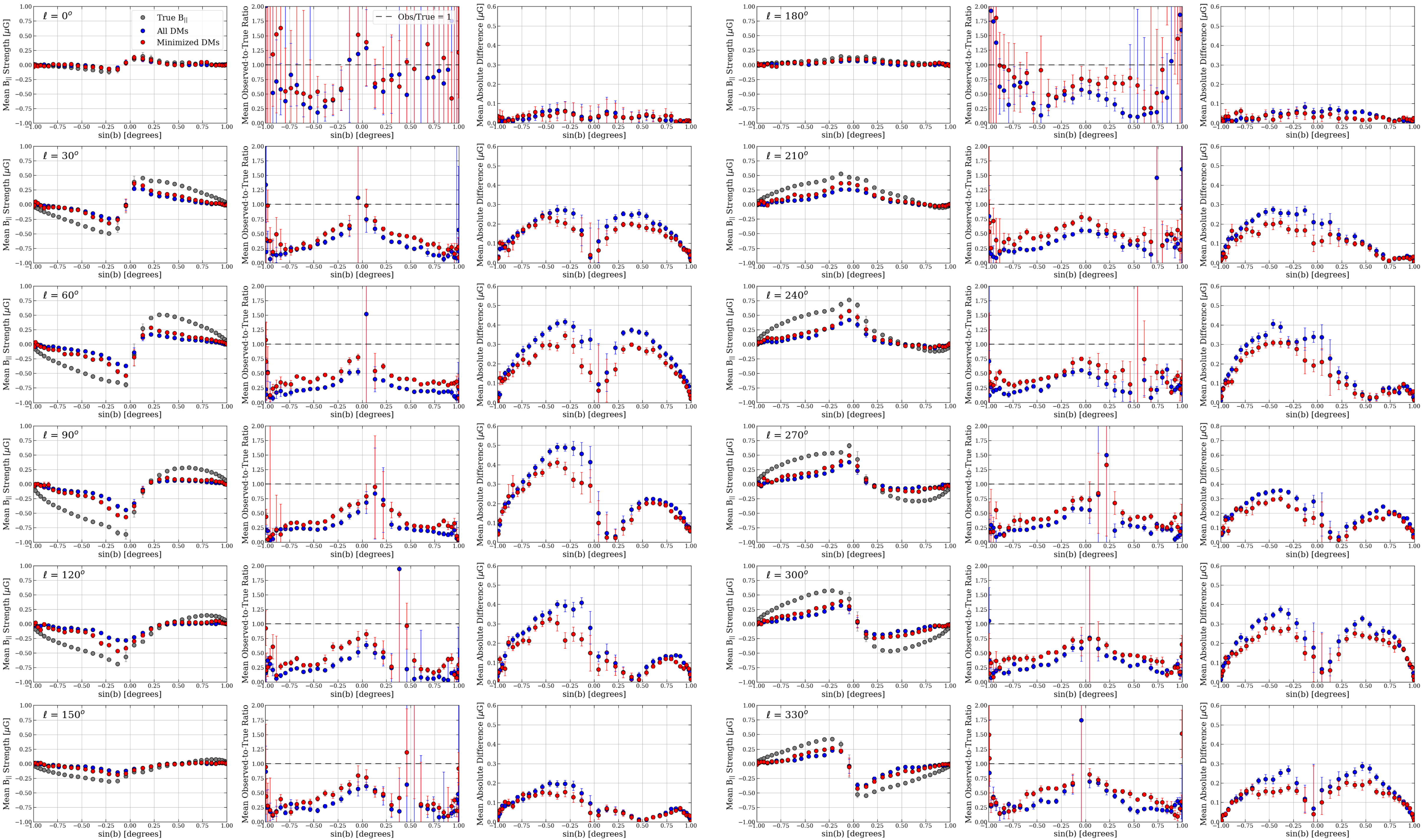}
    \caption{Same as Figure \ref{fig:bpar_analysis} but for the scenario without the X-shaped halo in the assumed GMF model.}
    \label{fig:bpar_noxhalo_analysis}
\end{figure*}

Figures \ref{fig:bpar_analysis} and \ref{fig:bpar_noxhalo_analysis} illustrate the offset between the expected $\left<B_{\parallel}\right>$ (grey points) and the reconstruction results (blue points for the full data set and red points for the DM filtered set), which primarily arises due to extragalactic DM contributions. Figures \ref{fig:bpar_analysis} considers the standard model (YMW16, YT20, JF12) used throughout the paper, while Figure \ref{fig:bpar_noxhalo_analysis} shows reconstructions after removing the X-shaped halo from JF12. Three sub-plots are produced for each slice centered at 30~deg intervals in Galactic longitude: (i) the first shows the mean $\left<B_{\parallel}\right>$ as a function of Galactic latitude; (ii) the second presents the observed-to-expected $\left<B_{\parallel}\right>$ ratio, with a dashed line representing where the ratio is unity; (iii) the third plots the absolute difference between the reconstruction results and the expected $\left<B_{\parallel}\right>$.

\begin{figure}
    \centering
    \includegraphics[scale=0.36]{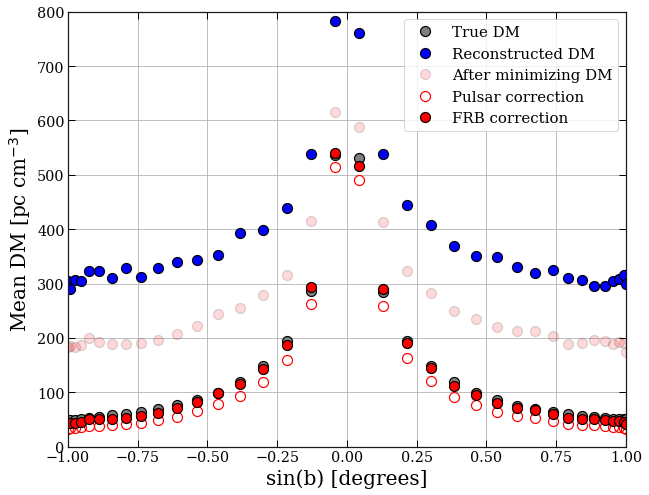}
    \caption{Analogous to Figure \ref{fig:corrfac} without the X-shaped GMF halo.}
    \label{fig:corrfac_noxhalo}
\end{figure}

\begin{figure}
    \centering
    \includegraphics[scale=0.195]{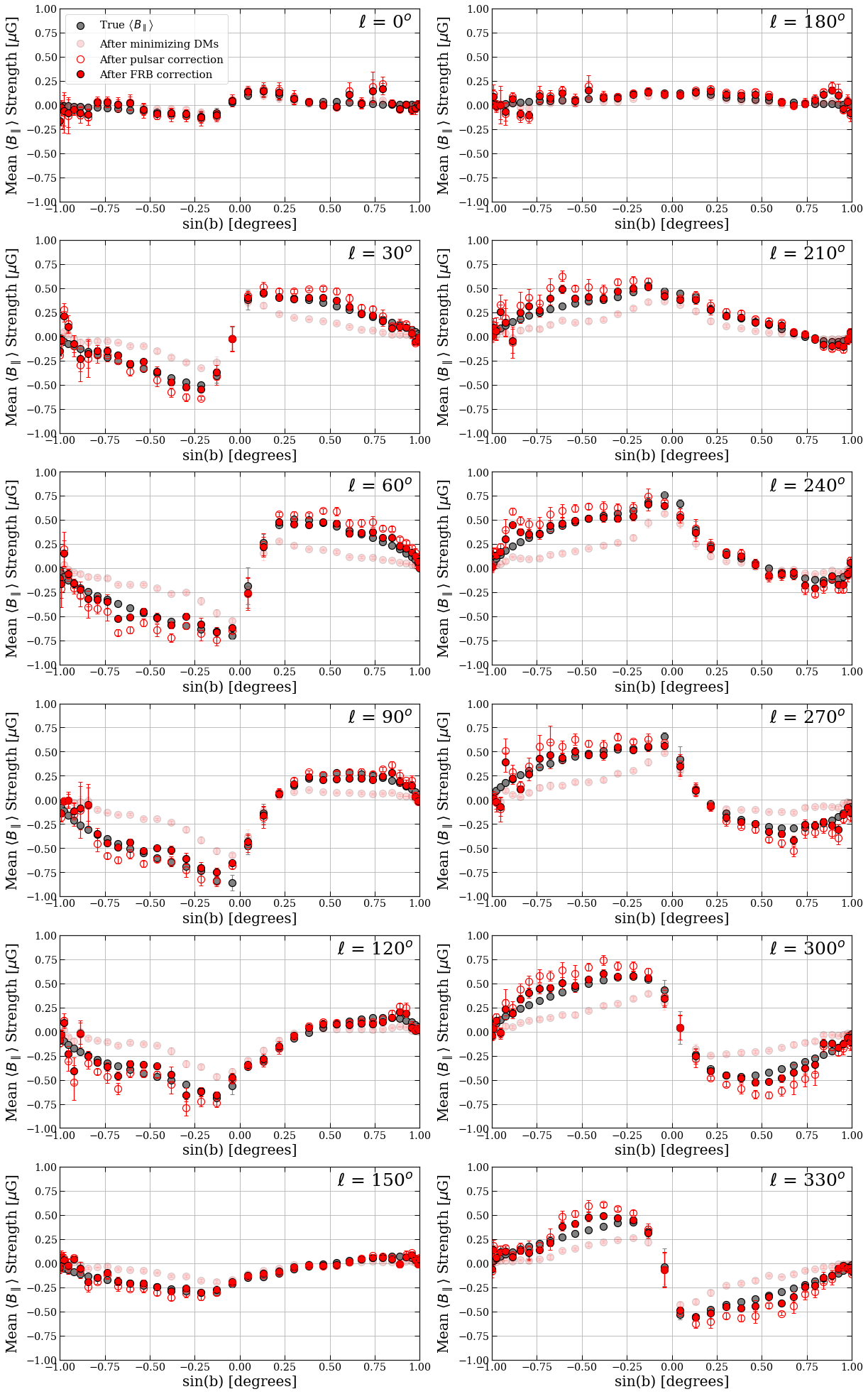}
    \caption{Analogous to Figure \ref{fig:corrfac_b} without the X-shaped GMF halo.}
    \label{fig:corrfac_long_noxhalo}
\end{figure}

Figures \ref{fig:corrfac_noxhalo} and \ref{fig:corrfac_long_noxhalo} are similar to Figures \ref{fig:corrfac} and \ref{fig:corrfac_b} in Section \ref{discussion} except after the removal of the X-shaped halo from the JF12 model. Again the same set of DM corrections in Section \ref{sec:dm_corr} (derived from equations \ref{eq:dm_model}$-$\ref{eq:dm_corr}) are applied to better recover Galactic structure.

\end{document}